# The long-term dynamics of

# co-authorship scientific networks:

# Iberoamerican countries (1973-2006)

Guillermo A. Lemarchand

Science, Technology and Society Graduate Program,
Universidad Nacional de Quilmes, Argentina

E-mail: lemar@correo.uba.ar

# The long-term dynamics of co-authorship scientific networks:

# Iberoamerican countries (1973-2006)

Guillermo A. Lemarchand[1]

Science, Technology and Society Graduate Program,

Universidad Nacional de Quilmes, ARGENTINA

**Abstract:** We study the national production of academic knowledge in all Iberoamerican and Caribbean countries between 1973 and 2007. We show that the total number of mainstream scientific publications listed in the Science Citation Index (SCI), the Social Science Citation Index (SSCI) and Arts and Humanities Citation Index (A&HCI) follows an exponential growth, the same as the national productivity expressed in the number of publications per capita. During the last 35 years, Portugal shows the highest rate growth in both indicators. We also explore the temporal evolution of the co-authorship patterns between a sample of 12 Iberoamerican countries responsible for 97.95% of the total regional publications between 1973 and 2006, with a group of other 45 different nations. We show that the scientific co-authorship among countries follows a power-law and behaves as a self-organizing scale-free network, where each country appears as a node and each co-publication as a link. We develop a mathematical model to study the temporal evolution of co-authorship networks, based on a preferential attachment strategy and we show that the number of co-publications among countries growths quadratically against time. We empirically determine the quadratic growth constants for 352 different co-authorship networks within the period 1973-2006. We corroborate that the connectivity of regional countries with larger scientific networks is growing faster than with other less connected countries. We determine the dates, $t_0$, at which the co-authorship connectivities trigger the self-organizing scale-free network for each of the 352 cases. We find that the last follows a normal distribution around year 1981.4 ±2.2 and we connect the last effect with a brain-drainage process generated during the previous decade. We show how the number of co-publications $P_k^i(t)$ between country $k$ and country $i$, is related with a power-law against the coupling growth coefficients $a_k^i$. We develop a methodology to use the empirically determined growth constants for each co-authorship network to predict changes in the relative intensity of cooperation among countries. We finally discuss the consequences of our findings for the science and technology cooperation policies.

## 1. Introduction

The application of bibliometric indicators to estimate the characteristics of international scientific cooperation patterns have been explored by diverse authors (i.e. Davison Frame et. al, 1977; Glänzel and Schubert, 2004; Holmgren and Schnitzer, 2004; Wagner and Leydesdorff, 2005b). Few studies employed them to analyze the cooperation profiles among Latin American countries (Fernández et al., 1998; Narvaes-Berthelemot, 1995; Russell, 1995; De Moya-Anegón and Herrero Solana, 1999; Gómez et al., 1999). Most of previous works only considered the aggregated behavior among short periods of time, no longer than a decade. By doing so, the information about the annual rate of change in cooperation networks is lost. In this way, no conclusions can be made about the results obtained by the application of different science and technology (S&T) international cooperation policies or the absence of them. The last might eventually be empirically quantified by contrasting the evolution of international cooperative agreements among institutions and countries against several S&T output indicators (Lemarchand, 2005). Unfortunately, this work was not carried out yet due to the lack of information about which international treaties and agreements are in operation among different countries.

[1] Professor at the Science, Technology and Society Graduate Program of the National University of Quilmes, Argentina. Postal Address: C.C. 8 – Sucursal 25, C1425ZAB Buenos Aires, Argentina. E-Mail: lemar@correo.uba.ar .

Besides the existence of several formal international cooperation instruments, there is no doubt that the informal ones, among individual scientists of different countries and disciplines, might explain most of the cases of co-authorship of scientific mainstream articles.

In this work, we determine the long-term evolution of the cooperation networks among 12 Iberoamerican and Caribbean countries[2] and other 45 regional and extra-regional nations. The selected countries are responsible for the 97.95% of the total main stream scientific publications written by scientists of this region that were listed in the Science Citation Index Extended (SCI), Social Science Citation Index (SSCI) and Arts and Humanities Citation Index (A&HCI), between 1973 and 2006 (34 years)[3]. We study the bilateral co-authorship of articles between each of the 12 countries with the 22 most productive Iberoamerican and Caribbean countries; 19 OECD countries not included in the first group and with China, India, Israel and USSR/Russia (see Table 2).

The analysis of the aggregated temporal evolution of SCI, SSCI and A&HCI, shows a homogeneous trend that is independent of any academic discipline and also avoids any substantial change in the national trends, due to the continuous incorporation of new journals to the databases. In this way, we focus our study in a cooperation network analysis within regional and extra-regional countries. Obviously, the publication in main-stream journals (listed by SCI, SSCI and A&HCI) represents only a fraction of all the cooperative research and development (R&D) activities that is taking place within the countries of our sample. The main advantage of using these databases is that they were systematically collected and organized over several decades with similar methodologies, allowing us to perform a long-term analysis with relative good confidence.

In Section 2 we analyze the long-term evolution in the production of mainstream scientific publications for all the countries of the region between 1973 and 2007, as well as their growth rates, productivities and regional distributions. In Section 3 we describe the methodology used to analyze the co-authorship patterns among a sample of 12 Iberoamerican countries and other 45 different nations between 1973 and 2006, and we present the main results. We study the co-authorship patterns in terms of intra and extra regional cooperation. We show that the intra-regional cooperation has been increasing smoothly during the last decades.

In Section 4 we develop a simple mathematical model of social networks applied to the study of the temporal evolution of co-authorship among countries. The model predicts a quadratic growth of co-publications (links) against time, among different countries (nodes). We show that this type of networks behaves with a self-organizing dynamics and we derive the conditions from which this process is triggered. In Section 5 we empirically analyze 352 different scientific co-authorship networks between 1973 and 2006 and we estimate the values of their growth constants, the dates at which the self-organizing dynamics starts and the correlation coefficients between the mathematical model and the real data. We determine the number of co-publications against the values of different growth coupling constants scales with a power-law. In Section 6, we use our mathematical model and the empirically determined growth constants, to deduce a methodology to predict the near-future behavior of the co-authorship patterns among the 352 collaborating networks. Finally, in Section 7, we present a summary with the main results of this research and their implications on the regional science and technology policies.

## 2. Iberoamerican mainstream knowledge production (1973-2007): the database

Within the most scientific productive 146 countries in all fields[4], covering a ten-year plus eight-month period (January 1997 to August 21, 2007), the only Iberoamerican countries included among the top twenty are: Spain (rank 10) and Brazil (rank 17). For the same period, our analysis shows that Mexico has the rank 27, while Argentina the rank 31. This is an interesting improvement[5] if we take into account that a similar survey made between 1967 and 1973, had Argentina (rank 27); Spain (rank 29), Brazil (rank 32), and Mexico (rank 34) as the most productive nations of the Iberoamerican and Caribbean region (De Solla Price, 1986: pp.192-193).

---

[2] The countries included here within the Iberoamerican and Caribbean region are: Antigua and Barbuda, Argentina, Barbados, Belize, Bermuda, Bolivia, Brazil, Chile, Colombia, Costa Rica, Cuba, Dominica, Dominican Republic, Ecuador, El Salvador, French Guiana, Guatemala, Guyana, Haiti, Honduras, Jamaica, Martinique, Mexico, Panama, Paraguay, Peru, Suriname, St. Lucia, St. Thomas, Trinidad and Tobago, Turks and Caicos, Uruguay, Venezuela, Portugal, and Spain. All the previous countries were taken into account to estimate the total number of regional publications per year (1973-2007). Here we have excluded Puerto Rico, because it is an associate State to the USA. Most of the small Caribbean islands have practically no mainstream scientific publications at all during this period.

[3] We tried to extend the search back to 1966, but according to the information provided by the technical support of Thomson-Reuters (MD-165137) on February 2007, the complete entries for authors, addresses and countries at the Web-of-Science, were only available for the SSCI since 1966, for SCI Expanded since 1973 and for A&HCI since 1975.

[4] http://www.in-cites.com/countries/2007allfields.html

[5] Here Argentina is an exception. Within this period there is a drop from rank 27 to rank 31.

In order to understand the process that took place between the last two extremes here we analyze the temporal evolution in the production of mainstream scientific and academic knowledge in all fields. In this Section all our analyses cover the period 1973 to 2007.

To study the distribution of published articles for each Iberoamerican and Caribbean country listed in SCI, SSCI and A&HC, we use Thomson-Reuters' Web-of-Science[6] as our information source. We think that these databases constitute a good and qualified indicator to inquire about knowledge-production patterns within the region. The data was downloaded in April 2007. Due to some delays in the publication of several journals and delays in the Web-of-Science data-entry, there is an approximately 10-12 % underestimation in the total number of published articles for the year 2007.

In this article we focus our study only in the production of mainstream scientific articles in all fields of knowledge. Over the years, several scholars debated about the underrepresentation of Third World journals in the last mainstream databases (Gaillard, 1991; Gibbs, 1995) and in particular the Latin American and Caribbean ones (Burgos, 1995). Not all the scientists of the region under study submit their research results to mainstream publications. Thus, the existence of domestic journals in several countries may reflect some peculiar domestic circumstances or specific scientific national agendas that are not considered by the mainstream databases. Other studies show that Latin American scientific journals don't have the minimum level of bibliographic control necessary to be uniquely identified, read and subscribe to, by an international audience (Cano, 1995). In this context, it is considered that periodical publications from such an infrastructure are condemned to a ghost-like existence, whereas the academics that publish in them will have their research results unrecognized. For these reasons, during the last decades, there has been an important increase in the submission of scientific articles, written by Iberoamerican & Caribbean authors, to mainstream publications.

In spite of these reluctant points of view[7] about the underrepresentation of local and regional journals, we argue that there is a good correspondence among SCI, SSCI, A&HCI and other international databases of scientific knowledge production. De Moya-Anegón and Herrero-Solana (1999) showed a strong correlation in the distribution of publications, for those articles written by Latin American authors, between the Science Citation Index Extended and other databases like PASCAL, INSPEC, COMPENDEX, CHEMICAL ABSTRACTS, BIOSIS, MEDLINE and CAB. They have obtained the following values for the correlation coefficient among these databases: $0.957 \leq R \leq 0.997$. This fact supports our hypothesis that the combination of SCI, SSCI and A&HCI might still be a good indicator for the study of mainstream scientific knowledge-production and the evolution of co-authorship networks among different countries.

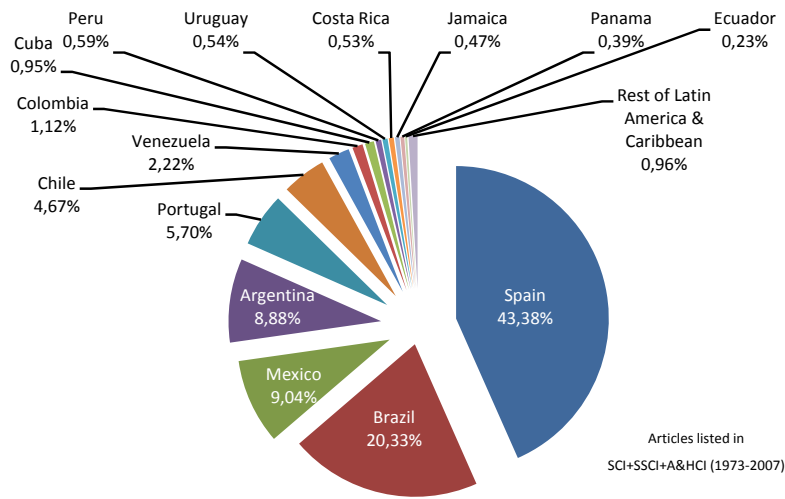

**Figure 1: Share distribution of mainstream scientific publications listed in the Science Citation Index (1973-2007), Social Science Citation Index (1973-2007) and Arts and Humanities Citation Index (1975-2007) for the Iberoamerican and Caribbean region. Here we represent those countries with more than 0.2 % of the total share. To estimate the total number of Iberoamerican and Caribbean regional publications we considered the following countries: Antigua and Barbuda, Argentina, Barbados, Belize, Bermuda, Bolivia, Brazil, Chile, Colombia, Costa Rica, Cuba, Dominica, Dominican Republic, Ecuador, El Salvador, French Guiana, Guatemala, Guyana, Haiti, Honduras, Jamaica, Martinique, Mexico, Panama, Paraguay, Peru, Suriname, St. Lucia, Trinidad and Tobago, Turks and Caicos, Uruguay, Venezuela, Portugal, and Spain. He we have excluded Puerto Rico, because it is an associate State to the USA.**

---

Figure 1 shows the distribution of published articles at the three databases, between 1973 and 2007, per country as a percentage of the total number of articles published in all the regional countries during the same period. Clearly, the European countries of our sample (Spain and Portugal), account for almost 50% of all the publications. This implies that if we want to calculate the shares of publications among Latin American and Caribbean countries alone, a rough estimation can be obtained by multiplying the values of Figure 1 by a factor of 2.

Figure 2 shows the total number of published articles at the three databases per year, between 1973 and 2007, in a log-linear scale for the 15 Iberoamerican nations with the highest number of mainstream scientific publications. The distribution of total number of publications per year, in most of the countries, follows an exponential growth behavior which can be described by $P_k(t) = \varphi e^{\gamma t}$, where $P_k(t)$ is the total number of publications of country $k$ at time $t$, while $\varphi$ and $\gamma$ are empirically determined constants. Table 1 shows the list of countries ordered by the total number of publications between 1973 and 2007, and by the size of the exponential growth constant $\gamma$. We also include the percentage shares and the exponential fitting correlation coefficient $R$ over 35 years of data.

It is very interesting to observe the difference between the total number publications and the exponential growth constant size rakings. Portugal has the highest $\gamma$ constant of our sample. In 1973 had the 7[th] position in the number of published articles of the region, while in 2007 got the 4[th] position, and now is almost sharing the 3[rd] one with Mexico that has 10 times Portugal's population. Another very interesting case is Colombia, which has the second highest $\gamma$-constant, with even a higher growth constant if we only consider the data after 1993. This is consistent with several S&T organizational reforms generated in 1990, such as the creation of the National Science and Technology System and the inclusion of COLCIENCIAS as part of the National Department of Planning (Garay et al., 1998). Ecuador with a very low production per year (≈10 articles per year in 1973) increased their mainstream publications in a factor of 20 over 35 years. Venezuela and Jamaica have the lowest $\gamma$ constant rates. The first one increased their mainstream publications in a factor less than 2 over 35 years, while the second practically has not changed its yearly production over the examined period.

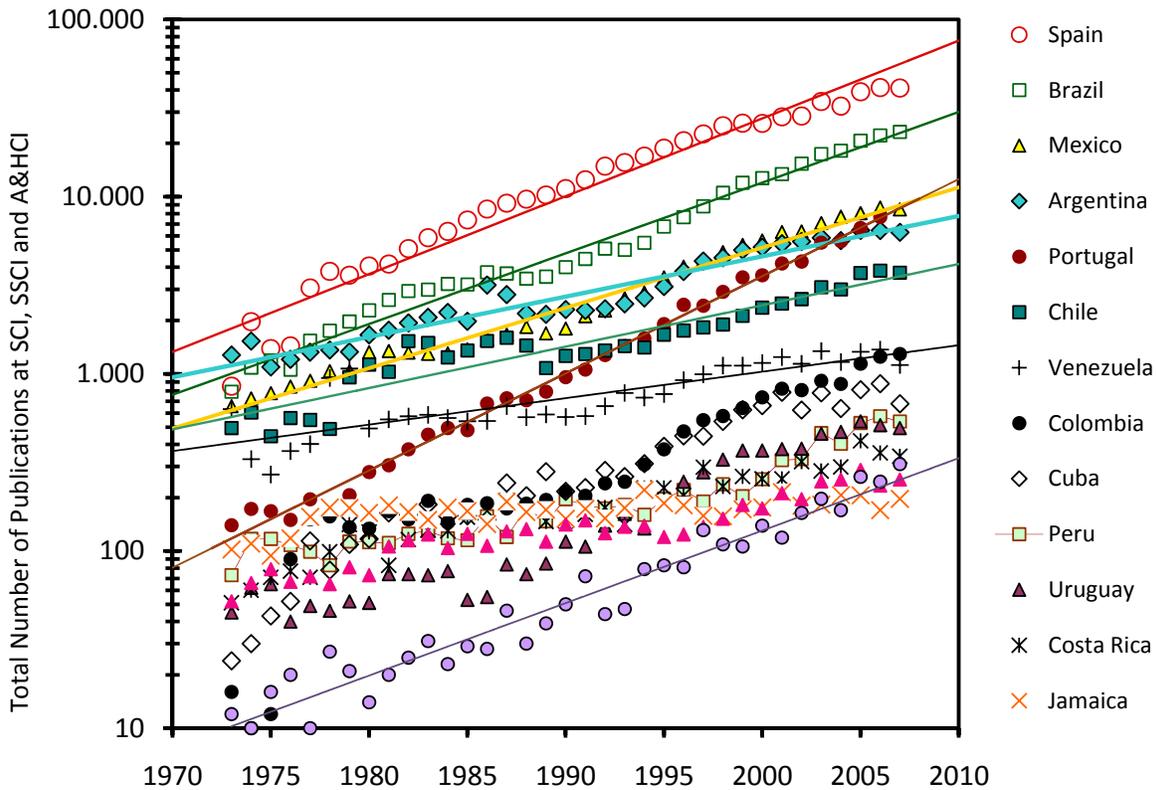

**Figure 2: Publication of peer-review articles listed in SCI, SSCI and A&HCI of the 15 most productive Iberoamerican and Caribbean countries (1973-2007). They approximately follow exponential growths. During this period, Portugal shows the highest growth rate.**

These bibliometric measurements belong to the S&T output indicators set (i.e. OECD Frascati Manual), and their study represents, in many ways, the performance of the national production of knowledge as a consequence of their R&D activities. The absence of growth, for more than three decades (i.e. Venezuela and Jamaica) implies a failure of the applied S&T policies in those countries. On the contrary, the S&T policies applied in Spain, Brazil and Portugal show a relative good success, at least in the production of mainstream scientific knowledge.

**Table 1: Iberoamerican Countries with the highest number of mainstream scientific publications between 1973 and 2006**

| N° | Ordered by the total number of publications between 1973-2006 | | | | Ordered by the growth constant | |
|---|---|---|---|---|---|---|
| | Country | % Share 1973-2006 | $\gamma$ | $R$ | Country | $\gamma$ |
| 1 | Spain | 43.38 | 0.1011 | 0.976 | Portugal | 0.1262 |
| 2 | Brazil | 20.33 | 0.0919 | 0.989 | Colombia | 0.1055 |
| 3 | Mexico | 9.04 | 0.0783 | 0.989 | Spain | 0.1011 |
| 4 | Argentina | 8.88 | 0.0525 | 0.964 | Ecuador | 0.0941 |
| 5 | Portugal | 5.70 | 0.1262 | 0.997 | Brazil | 0.0919 |
| 6 | Chile | 4.67 | 0.0537 | 0.931 | Cuba | 0.0901 |
| 7 | Venezuela | 2.22 | 0.0344 | 0.822 | Uruguay | 0.0821 |
| 8 | Colombia | 1.12 | 0.1055 | 0.906 | Mexico | 0.0783 |
| 9 | Cuba | 0.95 | 0.0901 | 0.958 | Chile | 0.0537 |
| 10 | Peru | 0.59 | 0.0499 | 0.935 | Argentina | 0.0525 |
| 11 | Uruguay | 0.54 | 0.0821 | 0.957 | Costa Rica | 0.0502 |
| 12 | Costa Rica | 0.53 | 0.0502 | 0.952 | Peru | 0.0499 |
| 13 | Jamaica | 0.47 | 0.0125 | 0.665 | Panama | 0.0402 |
| 14 | Panama | 0.39 | 0.0402 | 0.945 | Venezuela | 0.0344 |
| 15 | Ecuador | 0.23 | 0.0941 | 0.974 | Jamaica | 0.0125 |

Here $\gamma$ is the exponential growth constant and $R$ is the fitting correlation coefficient.

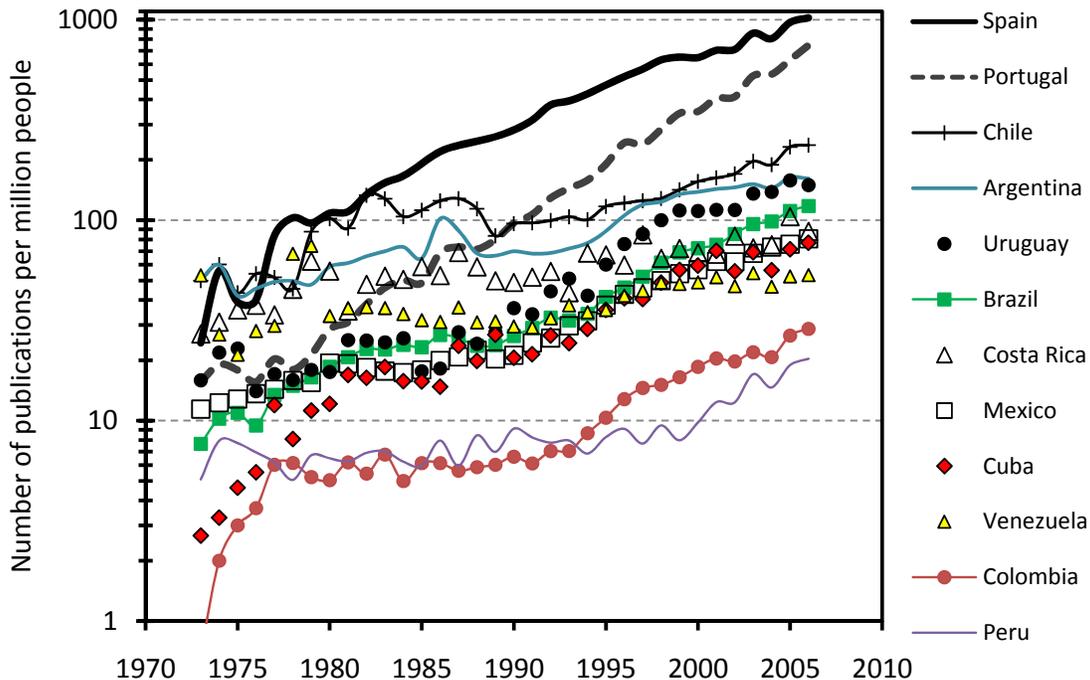

**Figure 3: Temporal evolution of total number of publications in the three databases (SCI, SSCI and A&HCI) per million inhabitants for each of the 12 Iberoamerican countries responsible for the 97.95% of the total number of articles published between 1973 and 2006 within the region.**

Another way to study the long-term behavior of mainstream scientific publications can be performed by analyzing the evolution of the societal knowledge productivity, in terms of number of publications per million inhabitants against time. In Figure 3 we show, in log-linear scale, the yearly distribution of publications at SCI, SSCI and A&HCI per million inhabitants, for the 12 most productive nations of the region. Both European countries (Spain and Portugal) show a clear exponential growth, in which Spain increased 43 times its productivity over 35-year period, while Portugal 47 times. By keeping the same growth rates, Portugal will reach Spain's productivity in the next decade. A careful look at the graph will also show that for most of Latin American countries there is a period, between the late seventies and the late eighties, when their productivity remained constant. A possible explanation might be the high brain-drain rate of scientists and technicians that the region suffered in this particular time due to the predominance of military governments (see Figure 18), economical crises, high inflation rates, the absence of academic freedom and explicit support to S&T activities.

From the mid-nineties, most of the nations of the region started increasing its productivity again. The exception is Venezuela whose productivity remained almost the same over 34 years (within a factor of ≈2). Again, the latter is also a good indicator of the failure of the applied S&T policies.

## 3. Iberoamerican mainstream co-publication patterns (1973-2006):

Co-authorship is one of the most tangible and well documented forms of scientific collaboration (Glänzel & Schubert, 2004). Almost every aspect of scientific cooperative networks can be studied by analyzing co-authorship patterns with the employment of bibliometric methods. For our study, we select only those nations with shares over 0.5 % of the total regional production between 1973 and 2006 (see Figure 1). In this way, the countries of our sample are: Spain, Brazil, Mexico, Argentina, Portugal, Chile, Venezuela, Colombia, Cuba, Peru, Uruguay and Costa Rica. This set covers the 97.95 % of the total number of publications originated in all Iberoamerican and Caribbean nations.

We also use the Web-of-Science resources to analyze the number of bilateral co-publications for each country of our sample, per year, between 1973 and 2006, with other 45 different nations taken from the list shown in Table 2. The last one includes the countries with larger number of mainstream scientific publications in the world, listed in SCI, SSCI and A&HCI[8]. The data were downloaded between February and March 2007. Again, due to some delays in the publication of several journals and delays in the Web-of-Science data-entry, there is an approximately 10-12 % underestimation in the total number of published articles for the year 2006, and consequently an underestimation in the number of co-publications between countries for that year.

We search for the number of publications per year (at the three databases) that have at least one author from the Iberoamerican country under study and another author from each of the 45 countries of our list. In this way, we got the number of bilateral co-publications that each pair of countries generates per year between 1973 and 2006. By knowing the total number of publications per year for each country, it is trivial to compute the co-publications as shares of the total annual production. We do not analyze how many of these bilateral co-authorship articles were really written by authors from three or more different countries. Our focus is the study of the links between countries. For doing so, we only need to know the existence of connectivity, or not, between each Iberoamerican country and each other nation from our list (Table 2).

Lemarchand (2007) showed that most of the co-publications are made with extra-regional countries. For example, he calculated that between 1966 and 2006[9], the average cooperation percentages for extra-regional *(E)* and intra-regional *(I)* co-authorships are respectively: Spain *(E≈37%, I≈5%)*; Brazil *(E≈34%, I≈7%)*; Mexico *(E≈66%, I≈7%)*; Argentina *(E≈27%, I≈13%)*; Portugal *(E≈63%, I≈11%)*; Chile *(E≈42%, I≈14%)*; Venezuela *(E≈35%, I≈13%)*; Colombia *(E≈67%, I≈33%)*; Cuba *(E≈28%, I≈28%)*; Peru *(E≈57%, I≈27%)*; Uruguay *(E≈41%, I≈35%)*; and Costa Rica *(E≈56%, I≈21%)*.

Figures 4 to 8 show the temporal evolution in the co-authorship for the five countries of our sample that have more than 5 % of the total number of publications of the region (1973-2006). All the co-authorship countries shown in these figures have more than 1.4 % of the total number of publications, published by each of these 5 nations.

During the period surveyed in our study, USA got the highest rates of co-authorship with the countries of the region: Spain≈7%; Brazil≈12%; Mexico≈46%; Argentina≈10%; Portugal≈10%; Chile≈15%; Venezuela≈16%; Colombia≈27%; Cuba≈4%; Peru≈28%; Uruguay≈15% and Costa Rica≈28%. The last figures are calculated taken into account the total number of co-publications with USA between 1973 and 2006, over the total number of publications of each Iberoamerican country within the same period.

---

[8] In this study: (1) the co-authorship patterns with United Kingdom (U.K.) are considered as a single unit. To estimate them, we have taken into account all those author's addresses located at England, Scotland, Wales and Northern Ireland; (2) in the case of Germany, between 1973 and 1989, we have considered as a single unit the aggregated co-publications of both: the German Democratic Republic (East Germany) and the Federal Republic of Germany (West Germany); (3) after 1992 the time series from the former USSR are followed by the Russian Federation data.

[9] Between 1966 and 1972 only the data for the Social Science Citation Index (SSCI) was available.

**Table 2: Countries of our sample**

| Countries Analyzed | Co-authorship with: | | |
|---|---|---|---|
| | Iberoamerican & Caribbean Countries | OECD Countries | Other Countries |
| a) Argentina | 1) Argentina | 24) Australia | 43) China |
| b) Brazil | 2) Barbados | 25) Austria | 44) India |
| c) Chile | 3) Bolivia | 26) Belgium | 45) Israel |
| d) Colombia | 4) Brazil | 27) Canada | 46) URSS/Russia |
| e) Costa Rica | 5) Chile | 28) Denmark | |
| f) Cuba | 6) Colombia | 29) Finland | |
| g) Mexico | 7) Costa Rica | 30) France | |
| h) Peru | 8) Cuba | 31) Germany | |
| i) Portugal | 9) Ecuador | 32) Ireland | |
| j) Spain | 10) Guatemala | 33) Italy | |
| k) Uruguay | 11) Guyana | 34) Japan | |
| l) Venezuela | 12) French Guiana | 35) Netherlands | |
| | 13) Honduras | 36) Norway | |
| | 14) Jamaica | 37) Poland | |
| | 15) Mexico | 38) South Korea | |
| | 16) Nicaragua | 39) Sweden | |
| | 17) Panama | 40) Switzerland | |
| | 18) Paraguay | 41) United Kingdom | |
| | 19) Peru | 42) United States | |
| | 20) Portugal | | |
| | 21) Spain | | |
| | 22) Uruguay | | |
| | 23) Venezuela | | |

An interesting case is Mexico, since the mid eighties, the co-authorship with USA has decreased from more than 60% of the total Mexican publications in 1985, to 40% in 2006. Brazil has been experiencing a similar decreasing co-authorship behavior with USA since 1995, while Portugal had an oscillatory pattern. The other 9 countries of our sample have been expanding their cooperation networks with USA over these years.

The rest of the most connected countries in terms of co-publications are Spain, UK, France, Germany, Brazil, Italy and Canada. From a scientific network dynamics point of view, the last countries (nodes) work like real co-authorship "hubs."

Lemarchand (2005) presented a short account of the history of multilateral cooperation, in science and technology, for Latin America and Caribbean countries since the late sixties. It was shown that the main multilateral programs were sponsored by the Organization of American States (OAS), UNESCO and since the mid-eighties by the Science and Technology for Development Iberoamerican Program (CYTED). On the other hand, there are several successful cooperation agreements and programs for specific thematic areas, among groups of countries. Examples of them are: the Latin American Biological Sciences Network (REDLAB), the Latin American Center of Physics (CLAF); the Argentinean-Brazilian Center for Biotechnology (CABBIO), the Andres Bello Agreement (Bolivia, Colombia, Chile, Ecuador, Panama, Peru, Spain, and Venezuela), MERCOSUR's Specialized Meeting in Science and Technology (RECYT), among other regional initiatives.

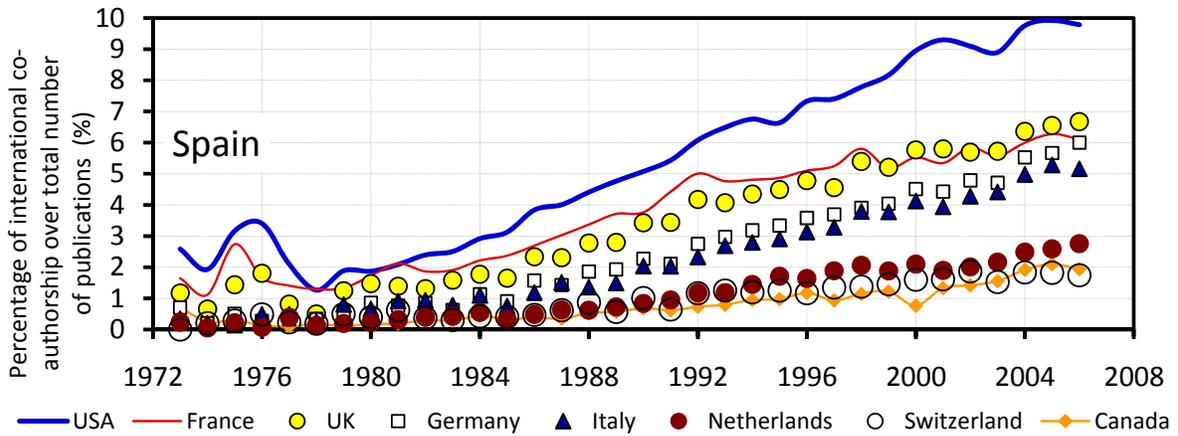

**Figure 4:** Temporal evolution of the percental distribution of international co-authorship of main-stream articles listed in SCI, SSCI and A&HCI for the 7 most important cooperative nodes (countries) responsible for co-publications with Spain. In the vertical axes we represent the number of co-publications as a percentage of the total number of documents published by the last, per year (1973-2006).

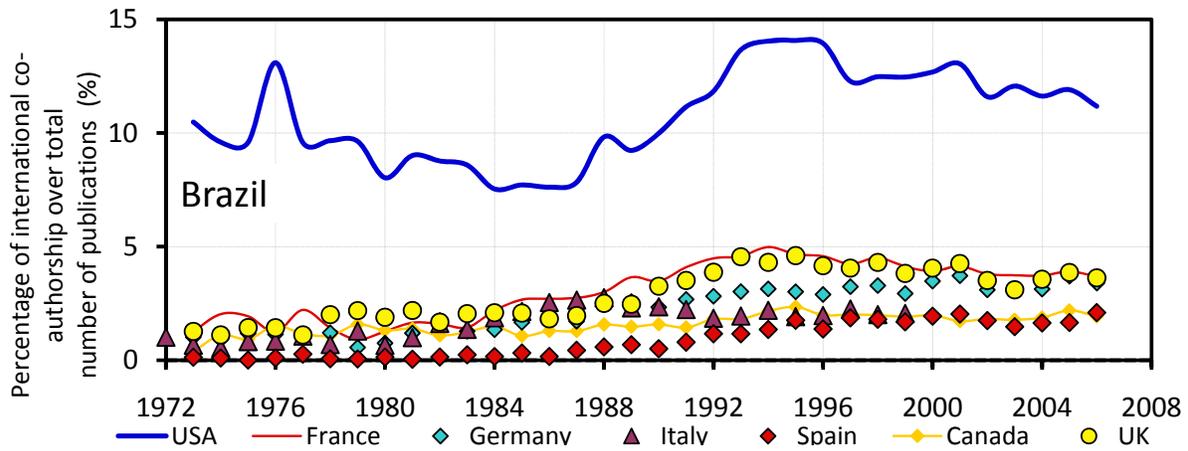

**Figure 5:** Temporal evolution of the percental distribution of international co-authorship of main-stream articles listed in SCI, SSCI and A&HCI for the 7 most important cooperative nodes (countries) responsible for co-publications with Brazil. In the vertical axes we represent the number of co-publications as a percentage of the total number of documents published by the last, per year (1973-2006).

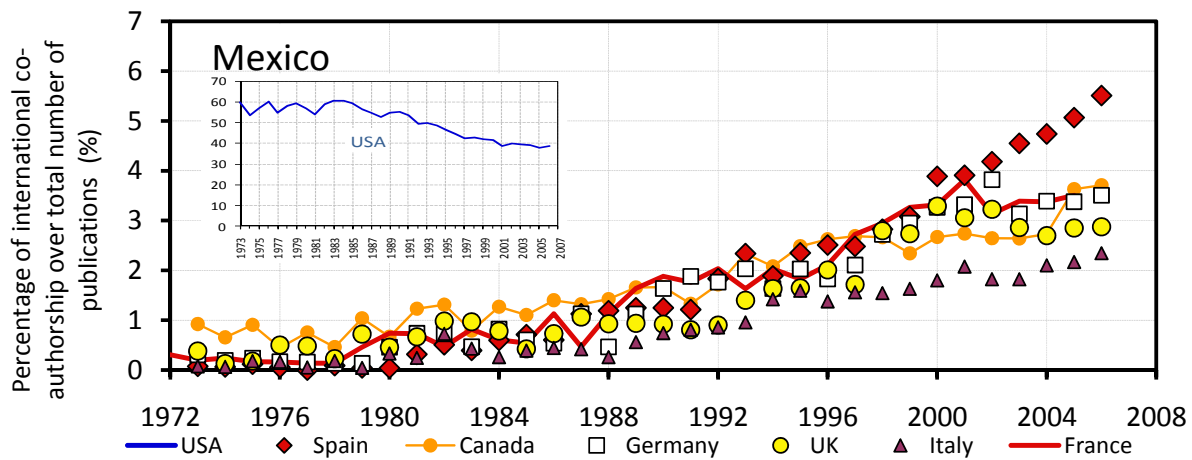

**Figure 6:** Temporal evolution of the percental distribution of international co-authorship of main-stream articles listed in SCI, SSCI and A&HCI for the 7 most important cooperative nodes (countries) responsible for co-publications with Mexico. In the vertical axes we represent the number of co-publications as a percentage of the total number of documents published by the last, per year (1973-2006).

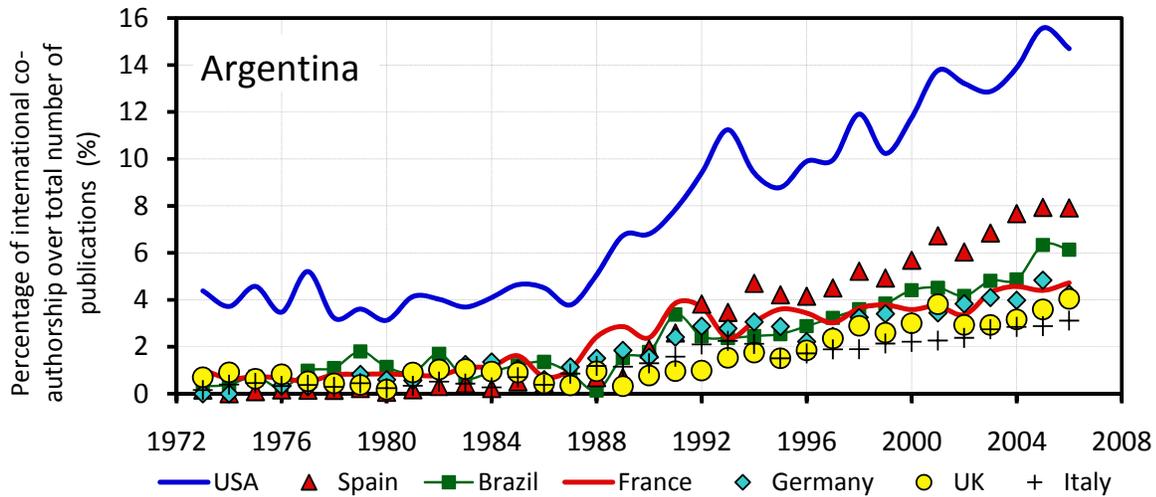

**Figure 7:** Temporal evolution of the percental distribution of international co-authorship of main-stream articles listed in SCI, SSCI and A&HCI for the 7 most important cooperative nodes (countries) responsible for co-publications with Argentina. In the vertical axes we represent the number of co-publications as a percentage of the total number of documents published by the last, per year (1973-2006).

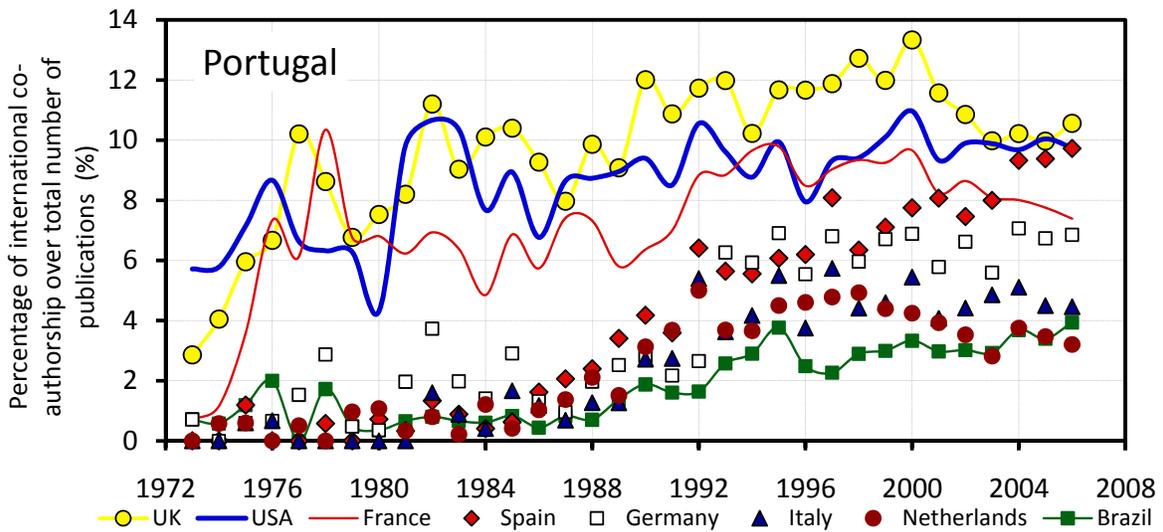

**Figure 8:** Temporal evolution of the percental distribution of international co-authorship of main-stream articles listed in SCI, SSCI and A&HCI for the 7 most important cooperative nodes (countries) responsible for co-publications with Portugal. In the vertical axes we represent the number of co-publications as a percentage of the total number of documents published by the last, per year (1973-2006).

It is not so clear which is the real impact of these regional programs on the production of new mainstream scientific knowledge. Using our database, we can roughly estimate the growth rate of internal collaboration among Iberoamerican and Caribbean countries. To analyze the temporal evolution of regional mainstream scientific co-publications we study the behavior of each nation of our sample with other 22 regional countries (99.83% whole mainstream publications) as an aggregated unit. In Figure 9 we show the internal co-publications growth, expressed as percentage over the total national number of publications. A careful analysis shows that the intra regional cooperation is increasing quadratically. At this point, we can not determine if this behavior is a consequence of the application of regional cooperation agreements and several explicit S&T policies driven by free-trade and other regional integration treaties, or not (Lemarchand, 2005). The number of international formal agreements has increased linearly in time over the last 40 years. It is also very interesting to check that the intensity of the aggregated cooperation with regional countries is inversely proportional to the size of the national scientific network (i.e. Spain is the country with the highest number of mainstream scientific articles, but the one with less regional cooperation shares). The last effect can be explained in terms of a self-organizing co-authorship scale-free network dynamics.

Albert and Barabási (2002) showed that highly connected nodes (in our case countries, i.e. USA, UK, France, Germany, etc.) increased their connectivity faster than their less connected peers. They called this effect as preferential attachment. Based on the last conceptual framework, Wagner and Leydesdorff (2005b) considered the growth of international scientific connectivities as a consequence of mechanisms of reputations and rewards, where scientists collaborate to gain visibility, reputation, access to state-of-the-art technologies and funding.

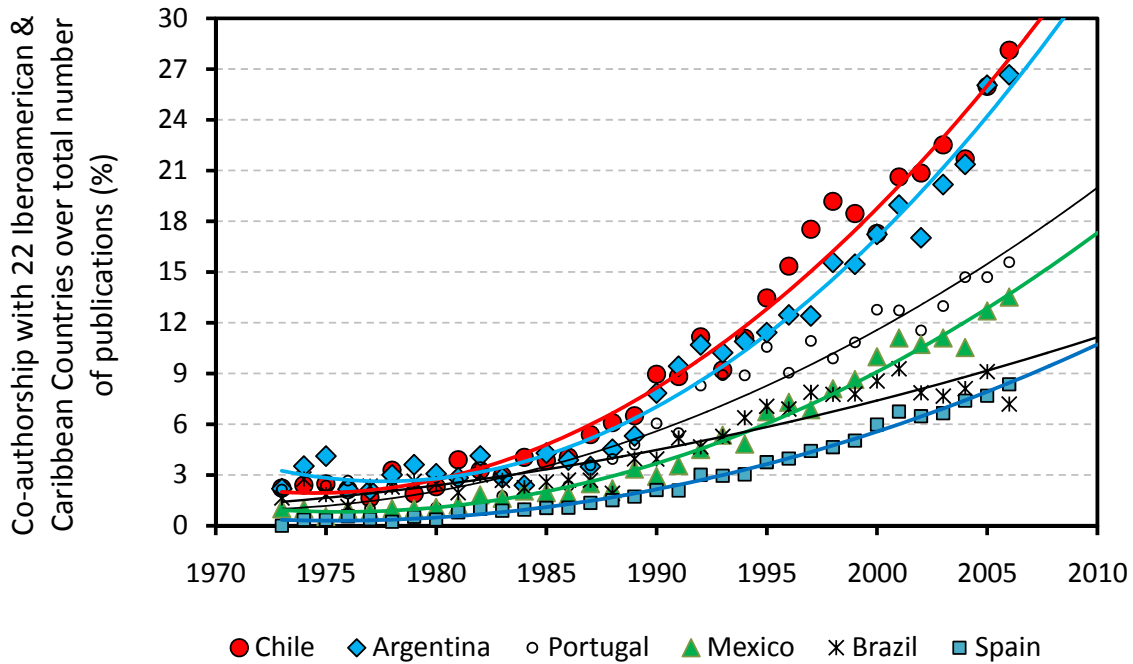

**Figure 9: Temporal evolution of the scientific co-authorship between Spain, Brazil, Mexico, Argentina, Portugal and Chile with the 22 most productive countries of the Iberoamerican and Caribbean region as a whole. The vertical axes represent the percentage of co-publications with Iberoamerican & Caribbean countries. Spain and Brazil work as real "hubs" where the rest of the regional countries concentrate most of their collaboration's shares with them, increasing in this way the internal percentage of co-authorship among Iberoamerican nations.**

It is interesting to check, that these main co-authorship links are established with nodes (countries) that have higher R+D budgets and extensive S&T networks than the Iberoamerican country under analysis. Assuming a complex network behavior, the strategy to extend a cooperative co-authorship strategy with countries that have a more extensive scientific network, is equivalent to the choice between holding a new document in a web server with high traffic (i.e. Google) or in a web-page with very few visits. In this way, scientists working on the periphery looking to increase the visibility of their research strive to link their research to the international research community, particularly through the co-publication with authors that are members of networks with larger connectivities (i.e. USA, UK, Germany, France, etc.).

From a sociological point of view, the individual scientists behave in such a way to enhance what Robert K. Merton (1968, 1988) called Matthew Effect[10]. It is very interesting to observe that Merton considered that *"the Matthew effect may serve to heighten the visibility of contributions to science by scientists of acknowledged standing and to reduce the visibility of contributions by authors who are less well known."* He also theorized that a macrosocial version of the Matthew Effect works as a maximization principle in those processes of social selection that currently lead to the concentration of scientific resources and talent. He had implicitly identified the sociological consequences of a power-law distribution applied to the prestige, access to resources, visibility of prominent scientists and institutions, or, in our terms, to the existence of preferential attachments.

One of the most common sources for establishing new links (preferential attachment) between a periphery country (node) and a "hub" is the international mobility of scientists, technicians, scholars and graduate students. This strategy was followed by individuals from most of the Iberoamerican countries over the last 50 years. The access to higher R&D budgets, better laboratories or infrastructures, and a scientific network with great connectivity and visibility, increase the individual productivity and give the possibility to collaborate later in hot topics with scientists of their own country. In this way, they extend the visibility of their national scientific network. This mobility is favored by international differences in earnings and technological gaps, the demand for talents from high developed countries, possibilities to work in the-state-of-the-art of the generation of new knowledge and new technologies. The last set of motivations can be considered as "pulling-factors."

---

[10] The Matthew effect took its name from the Gospel according to St. Matthew: "For unto every one that hath shall be given, and he shall have abundance: but from him that hath not shall be taken away even that which he hath." (Merton, 1968; p.58).

In turn, those "pushing-factors" which induce scientists, technicians, engineers and other scholars to emigrate are: low salaries at home, limited professional recognition, poor career prospects, and the absence of peer research groups at their home country (Solimano, 2008).

These highly skilled expatriate networks, through a connectionist approach, linking Diaspora members with their countries of origin, turn the brain drain into a brain gain approach (Meyer, 2001). In recent years, several empirical and theoretical studies that include the case of several Latin American countries support the last conceptions (Meyer et al, 1997; Kuznetsov & Sabel, 2006; Bassarsky, 2007; Solimano, 2008; Thron & Holm-Nielsen, 2008).

From the database generated by Lemarchand (2007) we empirically corroborated that the countries of the region (i.e. Figures 4 to 8) tend to mainly cooperate with "hubs" or bigger scientific networks. The higher number of articles is co-published with larger scientific networks. Table 3 shows the distribution of hubs and other larger co-authorships networks, the distribution of smaller collaborative webs and in both cases their corresponding percentage of co-publications over the total number of articles (1973-2006) for each country of our sample. As it was shown in Section 2, Spain has the rank 10 and Brazil the rank 17 within the most productive nations of the world. So, they respectively have only 9 and 16 larger scientific networks to cooperate with.

Table 3 shows the differences among the S&T cooperation strategies performed by the Iberoamerican countries of our sample. While several nations focus their co-authorship strategies on co-publications with larger networks (Colombia; Portugal, Peru, Uruguay and Mexico), others (Spain, Brazil and Argentina) share their mainstream scientific articles among local authors, hubs and smaller networks.

In Section 5 we study the characteristics of 352 different co-authorship Iberoamerican networks. The data provide in Table 3 shows that 70.2 % of all the networks surveyed in Section 5 are hubs or larger co-authorship networks. From the information of Table 3 and Figure 1 we estimate that the number of co-publications generated by the last group of networks represents 39.6 % of the total number of articles published by the 12 countries of our sample between 1973 and 2006. While the co-publications with smaller scientific co-authorship networks, is responsible for only 8.8 % of the total number of articles of these 12 nations. We notice that Spain (which works like a real "hub" for the Iberoamerican community) is the country responsible for contributing 6.1 % to the last 8.8 %.

**Table 3: Size distributions of Iberoamerican co-authorship networks**

| Iberoamerican Country | Number of larger co-authorship networks or "hubs" | Percentage of co-publications with larger networks [%] | Number of smaller co-authorship networks | Percentage of co-publications with smaller networks [%] | *Traffic Coefficient:* [Percentage larger networks/ Percentage smaller networks] |
|---|---|---|---|---|---|
| Spain | 9 | 28.37 | 30 | 13.76 | 2.1 |
| Brazil | 16 | 32.60 | 24 | 8.77 | 3.7 |
| Mexico | 26 | 69.52 | 14 | 3.25 | 21.4 |
| Argentina | 27 | 35.50 | 13 | 3.69 | 9.6 |
| Portugal | 26 | 73.97 | 5 | 0.82 | 90.2 |
| Chile | 28 | 53.30 | 7 | 2.23 | 23.9 |
| Venezuela | 21 | 44.49 | 4 | 1.68 | 26.5 |
| Colombia | 22 | 90.84 | 3 | 3.20 | 28.4 |
| Cuba | 22 | 53.29 | -- | -- | -- |
| Peru | 15 | 71.30 | 2 | 1.93 | 36.9 |
| Uruguay | 19 | 69.91 | -- | -- | -- |
| Costa Rica | 16 | 66.02 | 3 | 3.34 | 19.8 |

In the last column of Table 3 we represent the "traffic coefficient" or the ratio between the percentages of co-publications with larger scientific networks over the percentages of co-publications with smaller ones. Portugal's traffic coefficient is 90.2, several times larger than any other case of our sample (i.e. Spain: 2.1; Brazil: 3.7; Mexico: 21.4; Argentina 9.6, Chile: 23.9). By concentrating 90.2 times more co-publications with more visible and larger scientific networks than with smaller ones, the S&T cooperation strategy followed by Portugal allowed them to have the highest growth rates of mainstream scientific articles production and its corresponding productivity (see Figures 2 and 3).

### 4. The temporal evolution of self-organized co-authorship social networks:

The study of all kind of complex networks has undergone an accelerated expansion in the last few years, after the introduction of scale-free and power-law mathematical models (Albert & Barabási, 2002; Dorogovtsevyz & Mendes, 2002; Wang, 2002; Boccaletti et al. 2006) and small-world networks (Newman, 2001), which, in turn, have induced the study of many different phenomena under these new theoretical approaches. The co-authorship network is one of them (Barabási et al. 2002; Ramasco et al. 2004; Wagner & Leydesdorff, 2005a and 2005b; Tomassini, & Luthi, 2007). Nodes in co-authorship networks are paper authors, joined by edges if they have written at least one article together.

Ramasco et al (2004) also developed a mathematical model to show the self-organizing properties of all kind of collaboration networks. This model includes a growing network which combines preferential edge attachment with the bipartite structure and depends on the total number of collaborators (in our case countries) and acts of collaborations (in our case co-publications). According to this model, we can infer that co-authorship networks are self-organized and do not depend on any specific national S&T policy but on the internal dynamics of the scientific enterprise. Melin (2000) showed that the scientific collaborations are characterized by strong pragmatism and a high degree of self-organization. The same conclusions were treated with some detail by Wagner & Leydesdorff (2005a) suggesting that the scientific collaboration dynamics is caused by the self-interests of individual scientists, rather than to other structural, institutional or policy related factors that have been suggested previously.

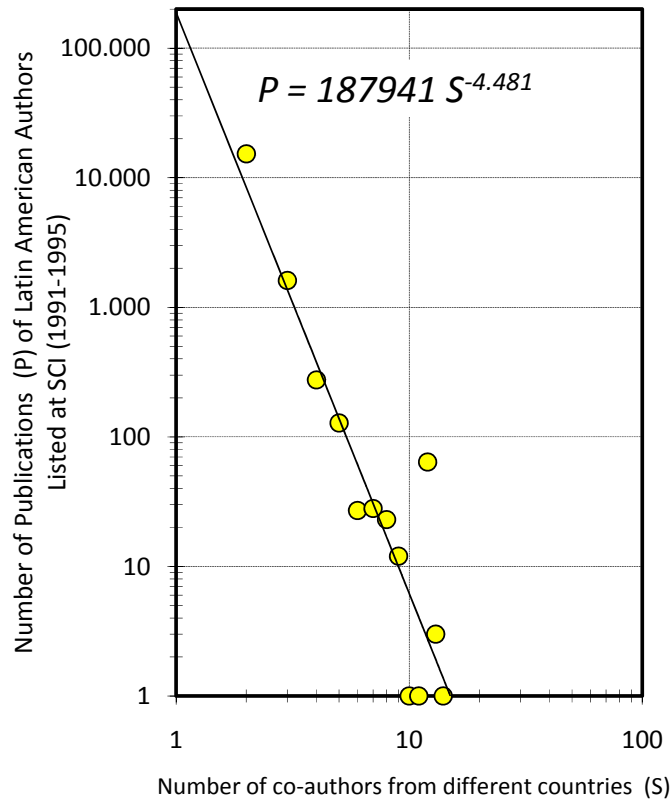

**Figure 10: Using the numerical data analyzed by Fernández et al. (1998), here we represent the distribution of mainstream scientific publications (links) by Latin American authors, listed in SCI between 1991 and 1995, against the number of authors from different countries (nodes). The graph shows a power-law distribution that is characteristic of a scale-free social network dynamics (Albert & Barabási, 2002).**

Using data provided Fernández et al. (1998) we verify that the co-authorship of Iberoamerican and Caribbean countries also behaves as a self-organizing social network. In Figure 10, we plot in a log-log scale the number of co-publications by Latin American authors (1991-1995) listed in SCI against the number of cooperative countries (i.e. $S = 2$; 3; 4; 14). The graph shows a power-law distribution, characteristic of a scale-free network (Albert & Barabási, 2002), where the countries behave as *nodes* and the co-publications as *links,* connecting nodes. The plotted data can be fitted by the following equation $P = 187941\ S^{-4.481}$, where $P$ is the total number of co-publications as a function of S (number of countries that participates in the co-publication). From the previous relation, it is clear that the most frequent type of co-authorship is the bilateral one, while the number of co-publications among 3; 4; ... 14 countries decays hyperbolically.

Besides the fact that several mathematical models have already been developed to understand the scientific co-authorship dynamics (Newman, 2001; Barabási et al. 2002; Ramasco et al. 2004; Li et al. 2007), here-on we develop a formalism to analyze the temporal evolution and the long-term behavior of co-publication networks among countries. Its simplicity will allow us to empirically determine its growth constants from available data-sets, to establish the dates at which the self-organizing network dynamics is triggered and to make some predictions about the future behaviors.

Using a continuum-theory network model (i.e. Barabási et al., 2002), here we denote $K_i(t)$ as the number of links that node $i$ has at time $t$; by $P(t)$ and $S(t)$ the total number of links (co-publications) and total number of nodes (scientists that are working at different countries) at time $t$, respectively. We consider that new researchers join the national scientific network at a constant rate, leading to:

$$S(t) = \sigma t + \theta \tag{1}$$

Here $\sigma$ and $\theta$ are constants. In this way, the average number of links per node in the system at time $t$ will be given by

$$\langle K \rangle = \frac{P(t)}{S(t)} \tag{2}$$

We assume that the probability to create a new internal link between two existing nodes is proportional with the product of their connectivities. By defining $\alpha$ as the number of newly created internal links per node in unit time, we are able to write the probability that between countries (nodes) $i$ and $j$, a new scientific paper (internal link) is published as:

$$\aleph_{ij} = \frac{2\alpha S(t) K_i K_j}{\sum_{l,m} K_l K_m} \tag{3}$$

The summation in the denominator is done for all values at which $l \neq m$. Assuming that a node $i$ has $K_i$ links, the probability that an incoming node will connect to it, is given by

$$\aleph_i = \beta \frac{K_i}{\sum_j K_j} \tag{4}$$

where $\beta$ is the average number of new links that an incoming node generates. Following Barabási et al. (2002) we are able to formulate the dynamical rules that govern our evolving network model, capturing the basic mechanism governing the evolution of co-authorship networks: (a) Nodes join the network at a constant rate; (b) Incoming nodes link to the already present nodes following preferential attachment; (c) Nodes already present in the network form new internal links following preferential attachment.

From (1) we know that new links join the system with a constant rate, $\frac{dS(t)}{dt} = \sigma$, the continuum equation that represents the behavior of the temporal evolution of the number of co-publications (links), associated with country (node) $i$, is the following master equation:

$$\frac{dK_i}{dt} = \sigma\beta \frac{K_i}{\sum_j K_j} + 2\alpha S(t) \sum_j \frac{K_i K_j}{\sum_{l,m} K_l K_m} \tag{5}$$

In the last equation, the first term on the right-hand side describes the contribution due to new nodes (4) and the second term gives the new links created with already existing nodes (3). The total number of links at time $t$ can be computed taking into account the internal and external preferential attachment rules:

$$\sum_i K_i = \int_0^t 2[S(t')\alpha + \sigma\beta]dt' = t\sigma(\alpha t + 2\beta) \tag{6}$$

Consequently, the average number of links per node increases linearly in time as:

$$\langle K \rangle = \alpha t + 2\beta \tag{7}$$

Combining equations (1), (2) and (7) we obtain:

$$P(t) = \langle K \rangle \, S(t) = (\alpha t + 2\beta)(\sigma t + \theta) \quad \text{or}$$

$$P(t) = \alpha \sigma t^2 + (\alpha \theta + 2\beta\sigma)t + 2\beta\theta \tag{8}$$

Equation (8) clearly shows the quadratic dependence of the number of co-publications between country $i$ and $j$, against time. By introducing a change of variables: $a = \alpha\sigma$; $b = (\alpha\theta + 2\beta\sigma)$; $c = 2\beta\theta$; we can re-write (8) as:

$$P(t) = at^2 + bt + c \tag{9}$$

By plotting the number of co-publications between country $i$ and country $j$, against time, and fitting the data with a quadratic equation, we are able to empirically obtain the values of the constants, *a, b* and *c*. In order to determine the time at which the self-organizing network starts working, we re-write (9) in a more convenient way, by introducing some algebraic transformations to the quadratic equation:

$$P(t) = a\left[t + \frac{b}{2a}\right]^2 + \frac{4ac - b^2}{4a} \tag{10}$$

For simplicity we introduce a new change of variables $t_0 = -\frac{b}{2a}$ and $\gamma = \frac{4ac - b^2}{4a}$, and we obtain the following relation:

$$P(t) = a[t - t_0]^2 + \gamma \tag{11}$$

In this way, the minimum of the function (11) will determine the value $t_0$ at which the self-organizing network dynamics is triggered. To find the value of $t_0$ we apply the following boundary conditions:

$$\frac{dP(t)}{dt} = 0 \rightarrow t = t_0 \tag{12}$$

and

$$\frac{d^2P(t)}{dt^2} = 2a \rightarrow \begin{cases} a > 0 \rightarrow Min \\ a < 0 \rightarrow Max \end{cases} \tag{13}$$

Consequently, the self-organizing co-publication network dynamics will start its process at $t_0 = -\frac{b}{2a}$ with $a > 0$. As it is shown in Tables 4 to 15 (see Appendix), all the coupling constants empirically obtained for the co-authorship networks between countries $k$ and $i$, are such that $a_k^i > 0$, which implies that for all our 352 different cases, $t_0$ is a minimum of the self-organizing social network.

## 5. The long-term empirical behavior of 352 bilateral co-authorship networks

In order to corroborate or falsify the model developed in Section 4, here we analyze the behavior of 540 different co-authorship networks, that result from studying the cooperation patterns among the 12 Iberoamerican countries of our sample and each of the 45 other nations listed in Table 2.

Assuming that each network will have at least two countries as nodes and several scientific publications as links which will increase quadratically against time, we analyze the number of publications that results from the cooperation of country $k$ with country $i$, as a function of time $P_k^i(t)$. From the analysis described in Section 3 we already have the quantity of articles that were co-published for 540 bilateral different networks between 1973 and 2006. Due to the fact that the quantity of mainstream scientific publications in diverse nations is very small, the co-publications among several pairs of countries practically do not exist, consequently, no network can be considered. From the original 540 bilateral co-publications study, in only 352 cases, it is possible to perform a network data analysis.

By fitting the data for each $P_k^i(t)$ distribution with quadratic equations, we can empirically determine the numerical values of the corresponding coefficients $a_k^i$, $b_k^i$, and $c_k^i$. From equations (10), (11) and (12) we calculate the value of $t_0$ at which the self-organizing network starts. In Tables 4 to 15 (see Appendix) we show the list of co-authorship networks among the 12 Iberoamerican countries of our sample with all those nations with enough data to perform a quadratic fitting.

These tables include information about the co-authorship country; the number of accumulated co-publications (1973-2006); the percentage $\left[\left(\frac{\sum_{j=1973}^{j=2006} P_k^i(t_j)}{\sum_{j=1973}^{j=2006} P_k(t_j)}\right) * 100\right]$ or the fraction of the total number publications between 1973 and 2006 of country $k$ that was co-published with country $i$; the coefficients $a_k^i$, $b_k^i$, and $c_k^i$; the time $t_0$ at which the self-organizing network is triggered; and the correlation coefficient (Weisberg, 1980) $R = \frac{cov(P(t_j), t_j)}{\sqrt{var(P(t_j)) var(t_j)}}$, here $j$ indicates years between 1973 and 2006.  Figures 11 to 15 show the co-authorship network behavior of Spain, Brazil, Mexico, Argentina and Portugal (Iberoamerican countries which account for 87.33% of the total mainstream scientific publications of the region). In these figures we represent the total number of co-publications (links) with their main cooperative nations (nodes) against time (1973-2006). It is self-evident how well the scale-free self-organizing network model fits the empirical data over 34 years. The minimum of these curves shows the value of $t_0$ at which the network dynamics is triggered behaving as a self-organizing process. According to this interpretation, the scientific collaborations for $t < t_0$ is below the threshold of connectivities needed to start the self-organizing process. For this reason, we also observe some sort of random behavior in the $P_k^i(t < t_0)$ values.

From the values of the growth coupling constants $a_k^i$, available at Tables 4 to 15, we can also verify the existence of the preferential attachment effect, where the connectivity of Iberoamerican countries with "hubs" or larger scientific networks (i.e. USA, UK, France, Germany, etc.) is growing faster than with other less connected countries (smaller scientific networks). The last is another prediction of the model that is corroborated by our empirical study.

Using the data from the 352 networks (see Appendix) we also empirically find that the total number of co-publications between countries $k$ and $i$, scales with a power-law with the coupling growth constant $a_k^i$ as $P_k^i = 5134.2\, a_k^{i\,0.9655}$ (see Figure 16). The last relation implies that the co-publications with "hubs" (i.e. USA, UK, France, Germany, etc.) growth hyperbolically much faster than with the rest of minor co-authorship networks. This fact is coherent with a preferential attachment strategy as it is predicted by our model.

When we analyze the distribution of the dates, $t_0$ at which the self-organizing network dynamics is triggered, among the 352 different networks, we find a normal distribution (see Figure 17) with the most probable value at $t_0 = 1981.4 \pm 2.2$. The first obvious conclusion indicates that the appearance of Internet was not a key issue for the emergence of co-authorship scientific networks dynamics. According to Laudel (2001), most of the scientific collaborations begin face-to-face. Historically relationships, former colonial ties and geographic proximity can only account for the internal Iberoamerican collaboration links (i.e. most of the cooperation with smaller co-authorship networks), but not for the co-authorship dynamics with "hubs". For the last, a preferential attachment generated by a brain-drain process is still the best explanation. As it was shown previously, there is an increasing set of new evidence showing that migration and mobility of researchers and scientists towards the mainstream scientific countries (hubs), expands the connectivities between the original country and a most visible and larger scientific network. This is probably the mechanism that triggers this scientific network dynamics.

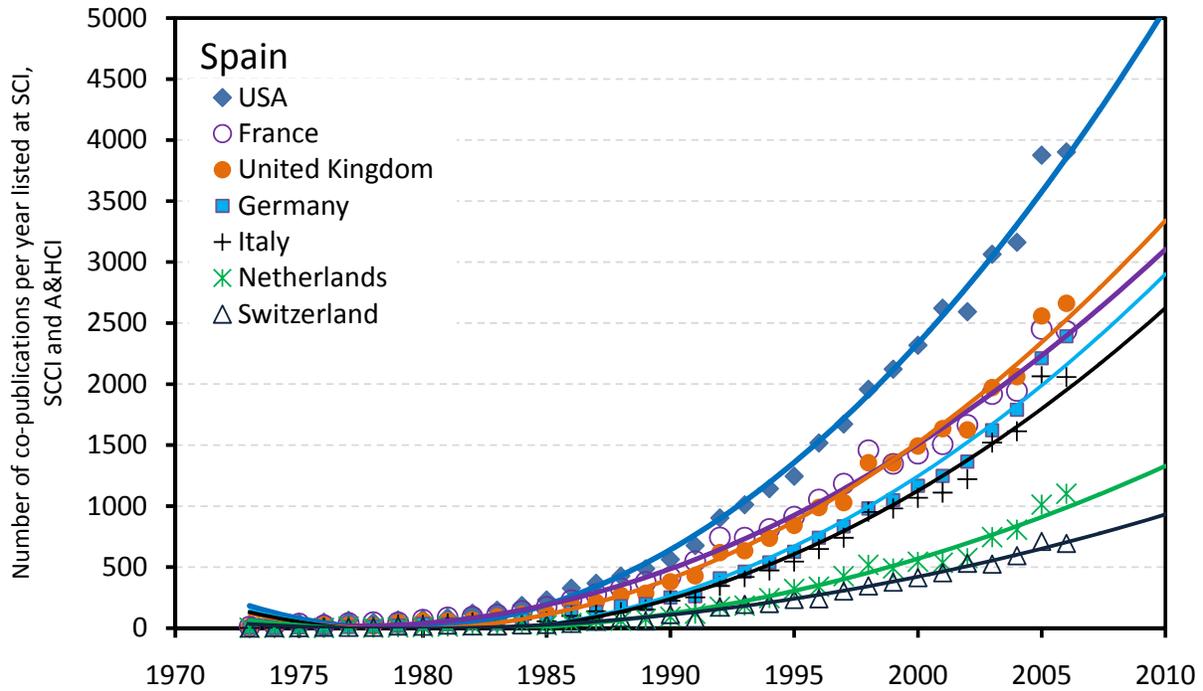

**Figure 11:** Temporal evolution of the co-authorship social network of Spain. Here we represent the number of co-publications against time for the 7 most important cooperative nodes (countries). The model developed predicts a parabolic growth in the number of links (publications) against time. Here the solid lines represent the quadratic fitting according to our model.

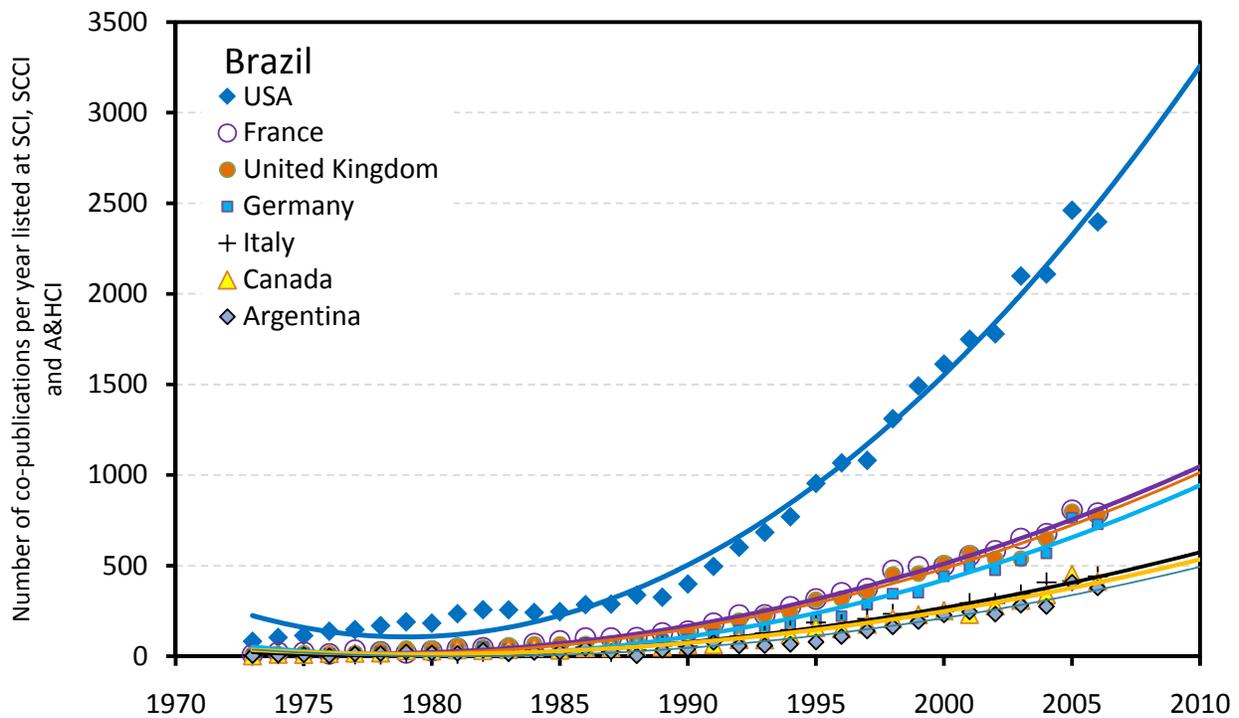

**Figure 12:** Temporal evolution of the co-authorship social network of Brazil. Here we represent the number of co-publications against time for the 7 most important cooperative nodes (countries). The model developed predicts a parabolic growth in the number of links (publications) against time. Here the solid lines represent the quadratic fitting according to our model.

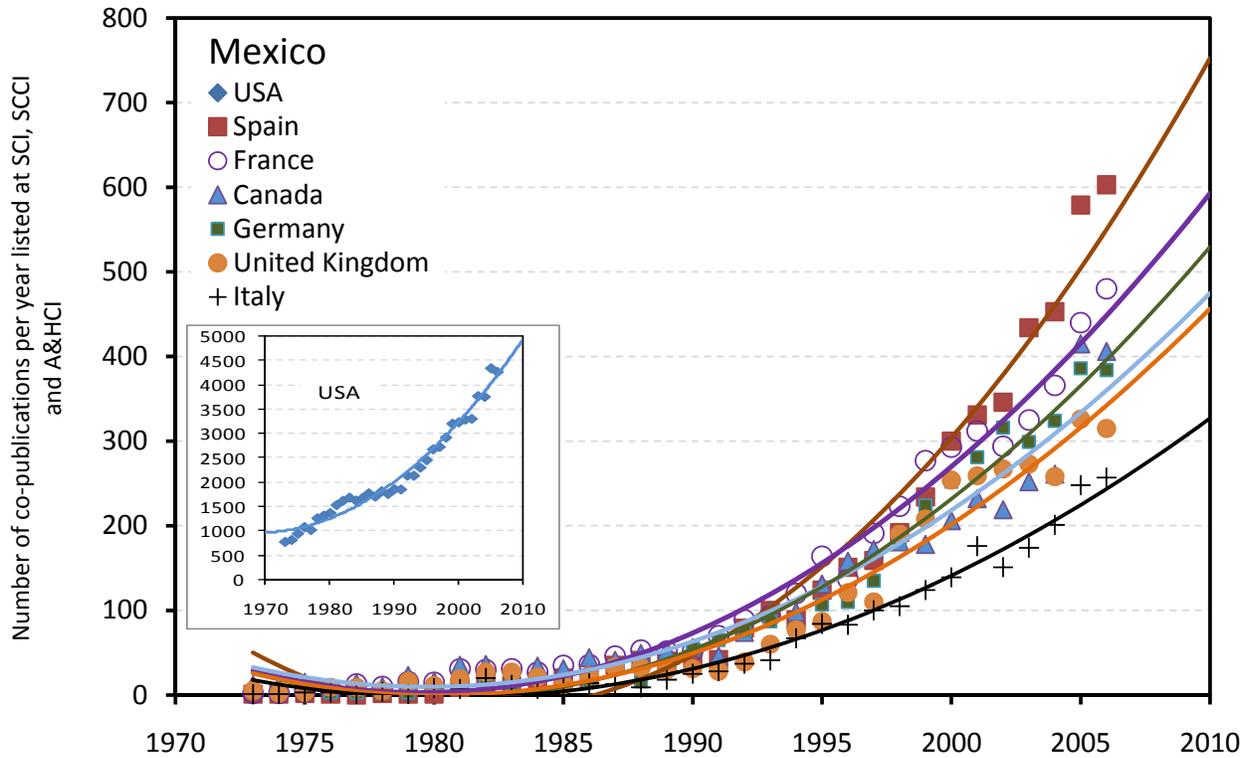

Figure 13: Temporal evolution of the co-authorship social network of Mexico. Here we represent the number of co-publications against time for the 7 most important cooperative nodes (countries). The model developed predicts a parabolic growth in the number of links (publications) against time. Here the solid lines represent the quadratic fitting according to our model.

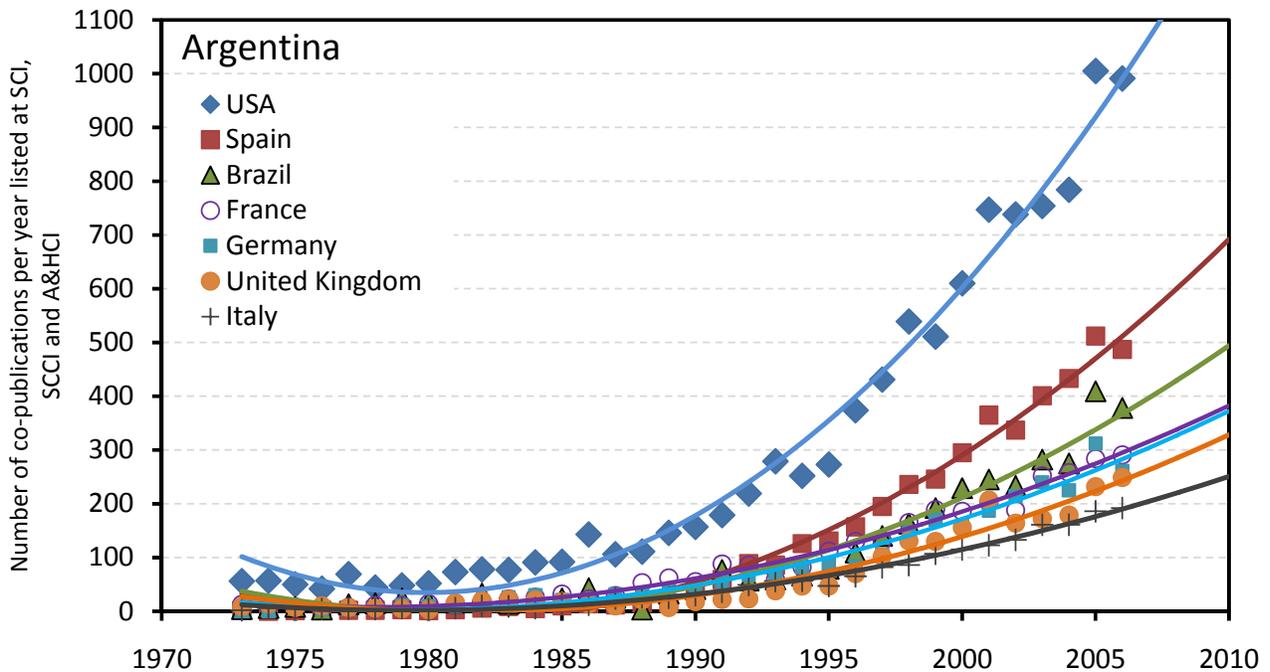

Figure 14: Temporal evolution of the co-authorship social network of Argentina. Here we represent the number of co-publications against time for the 7 most important cooperative nodes (countries). The model developed predicts a parabolic growth in the number of links (publications) against time. Here the solid lines represent the quadratic fitting according to our model.

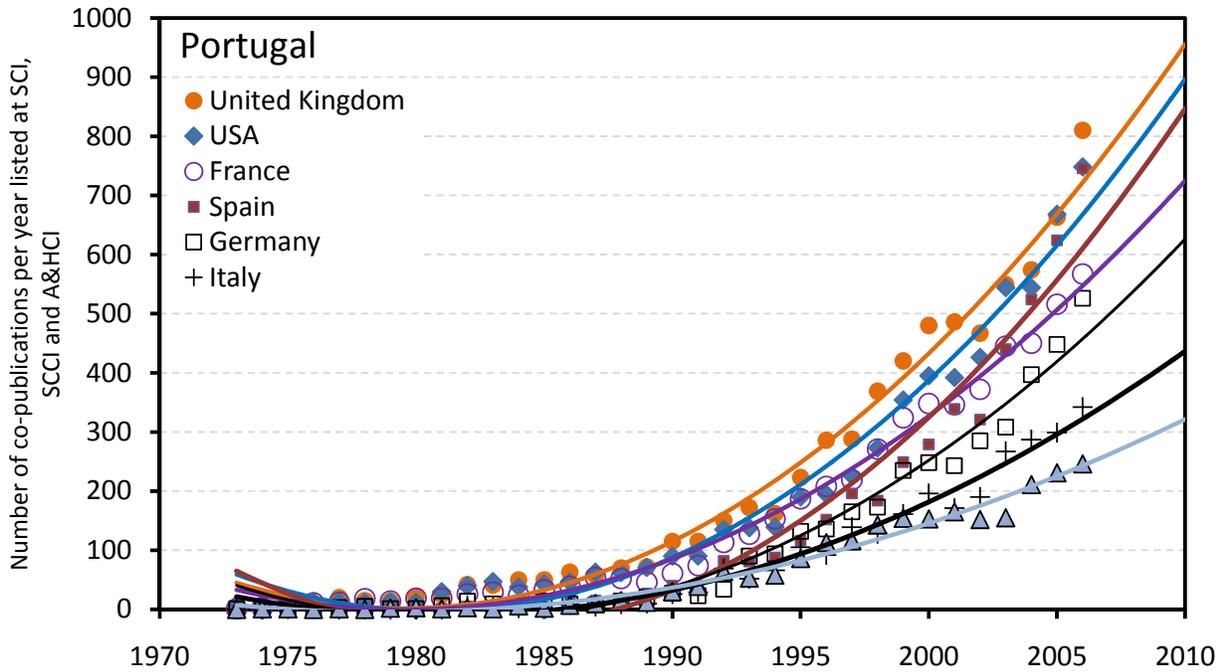

**Figure 15:** Temporal evolution of the co-authorship social network of Portugal. Here we represent the number of co-publications against time for the 7 most important cooperative nodes (countries). The model developed predicts a parabolic growth in the number of links (publications) against time. Here the solid lines represent the quadratic fitting according to our model.

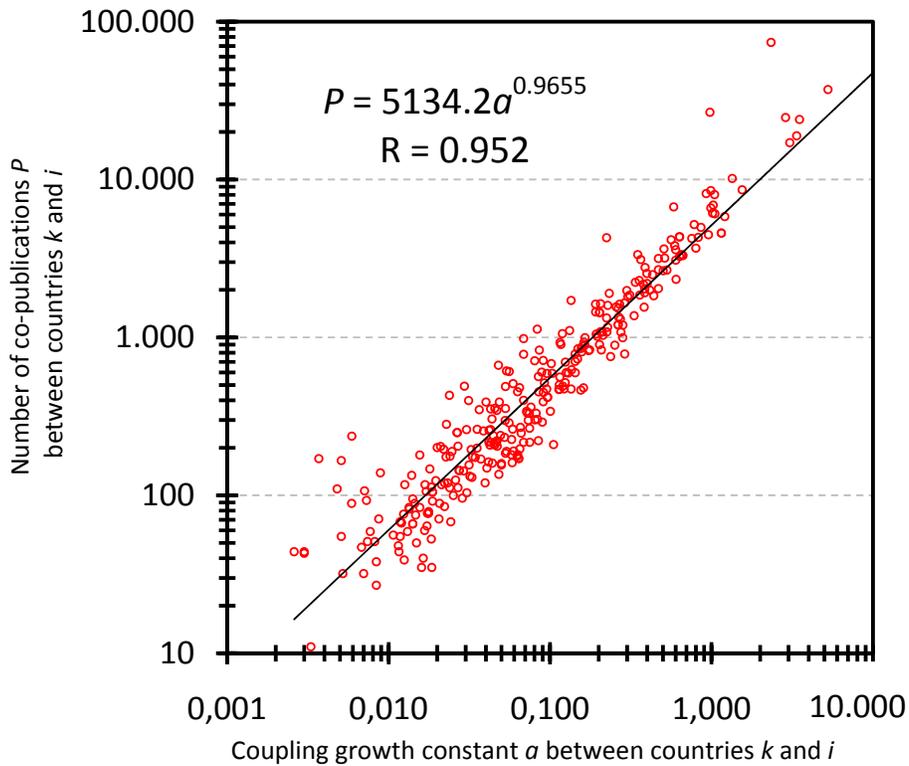

**Figure 16:** Power-law (scale-free) distribution total number of co-publications $P_k^i$, for each of the 352 co-authorship social networks, against the coupling constants $a_k^i$, according to the data taken from Tables 4 to 15.

Based on relations (10), (11) and (12), we know that there is a lower threshold of connectivities needed to trigger the self-organizing co-authorship network between two countries. According to the normal distribution shown in Figure 17, most of the co-authorship networks in Iberoamerican countries started within a short period of time (1981.4 ± 2.2). This effect might be explained by the massive brain-drain movement towards highly developed countries (Oteiza, 1971). In many cases, this emigration was generated by the adverse political situation among most of Iberoamerican nations, between the late sixties and mid eighties.

To show the severity of the last situation, in Figure 18 we estimate the percentage of the whole Iberoamerican and Caribbean's population, which were living under dictatorship's governments (1960-2006). The continuous line represents $\left(\frac{\sum_j I_j^D(t)}{\sum_j I_j(t)} \cdot 100\right)$, where $I_j^D(t)$ indicates the population as a function of time (1960-2006) of countries $j$ that had a dictatorship government $D$ at time $t$, and $\sum_j I_j(t)$ the temporal evolution of the population of all the countries $j$, that in this case are all those listed in the Footnote 2 (whole Iberoamerican and Caribbean population). The continuous line with squares, represents $\left(\frac{\sum_k I_k^D(t)}{\sum_k I_k(t)} \cdot 100\right)$, where $I_k^D(t)$ indicates the population as a function of time (1960-2006) of country $k$ from our sample of 12 Iberoamerican nations, that had a dictatorship government $D$ at time $t$, and $\sum_k I_k(t)$ the temporal evolution of the populations of all the countries from our sample.

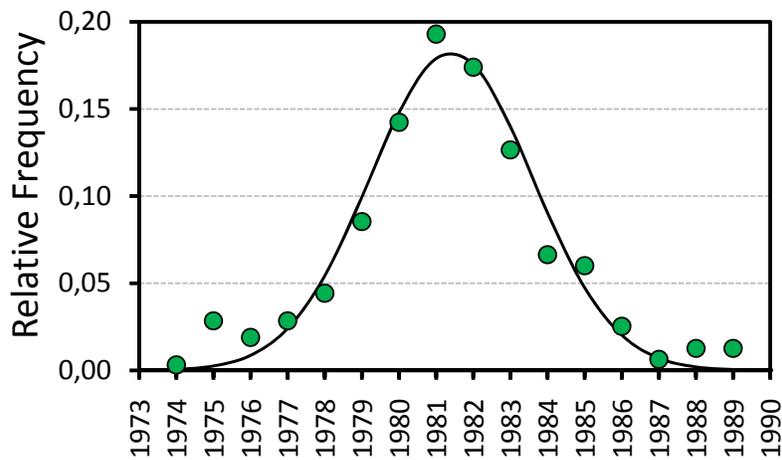

Figure 17: The dates *(t₀)* at which the co-authorships social networks start working as a self-organizing process, for the 352 scientific networks (see Tables 4 to 15) follows a normal distribution around year 1981.4 ± 2.2.

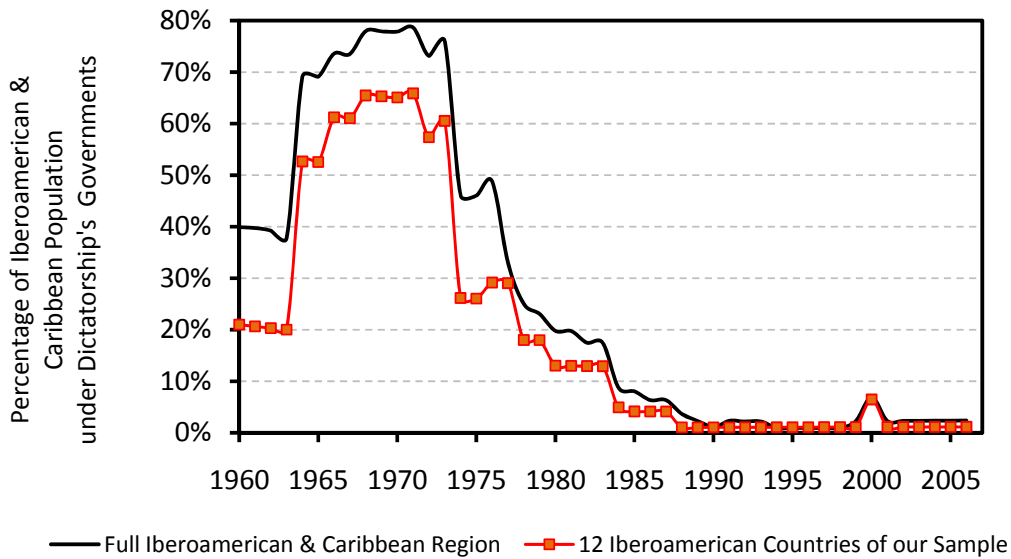

Figure 18: Percentage of the whole Iberoamerican and Caribbean countries population, living with Dictatorship's Governments, against time. We also represent the evolution of the Dictatorship's Governments among the 12 Iberoamerican countries of our sample expressed as a percentage of their whole population

Between the mid-sixties and mid-eighties, the Iberoamerican Region was dominated by a very difficult and delicate political and economical situation, which expelled most of their highly qualify talents (in science, technology, medicine, arts, literature and culture in general). The absence of academic freedom, persistence of ideological harassment, unstable macro-economical performance, low research budgets and wages, erratic S&T policies, constituted an extremely powerful driving-force for emigration (pushing-factors). These effects were also amplified by those pulling-factors described in Section 3.

After living for some time in developed nations, and participating in larger R&D networks, those expatriate scientists, engineers and technicians generated a critical mass of connectivities that trigged the co-authorship network dynamics, linking Diaspora members with scientists in their original countries.

## 6. Forecasting methods to study the scientific co-publication patterns among countries:

If the co-publication at time $t$ of country $k$ with two countries $i$ and $j$, have coupling coefficients $a_k^i > a_k^j$, we will obtain $P_k^i(t) > P_k^j(t)$ for every $t > t_0$. But if the coupling coefficients have the property that $a_k^i < a_k^j$, we can estimate the time $t_{int}$ at which both countries $i$ and $j$ have the same number of co-publications with country $k$ or $P_k^i(t_{int}) = P_k^j(t_{int})$. After this time ($t_f > t_{int}$) the number of co-publications of country $k$ with countries $i$ and $j$ will have the following opposite relation: $P_k^i(t_f) < P_k^j(t_f)$.

Here-on we estimate the value of $t_{int}$ as a function of the values $a_k^i, a_k^j, b_0^i, b_0^j, c_k^i, c_k^j$ that are empirically determined and listed in Tables 4 to 15 (see Appendix).

If

$$P_k^i(t_{int}) = P_k^j(t_{int}) \tag{14}$$

From (9) and (14) we have

$$a_k^i t_{int}^2 + b_k^i t_{int} + c_k^i = a_k^j t_{int}^2 + b_k^j t_{int} + c_k^j \tag{15}$$

and

$$0 = \left(a_k^j - a_k^i\right)t_{int}^2 + \left(b_k^j - b_k^i\right)t_{int} + \left(c_k^j - c_k^i\right) \tag{16}$$

Introducing the following changes of variables:

$$\begin{cases} \mu = \left(a_k^j - a_k^i\right); \\ \rho = \left(b_k^j - b_k^i\right); \\ \omega = \left(c_k^j - c_k^i\right); \end{cases} \tag{17}$$

We finally can determine the value of $t_{int}$ as:

$$t_{int} = \frac{-\rho + \sqrt{\rho^2 - 4\mu\omega}}{2\mu} \tag{18}$$

Here we need to consider only the major real root of equation (18) that corresponds to the moment at which a smaller co-authorship network ($P_k^i(t) > P_k^j(t)$) but with a higher growth coupling coefficient, $a_k^i < a_k^j$, reaches the same annual co-publications rate than the biggest one ($P_k^i(t) = P_k^j(t)$).

We apply this methodology to a particular substitution case to show how it works. For many years, the second main co-authorship country of Spain was France, but recently the last was replaced by the United Kingdom (see Figure 11). Here we apply our formalism to estimate $t_{int}$ by using the coefficients taken from Table 4: $a_{Spain}^{France} = 2.8737$, $b_{Spain}^{France} = -11364.005$, $c_{Spain}^{France} = 11234641.97$, $a_{Spain}^{UK} = 3.5101$, $b_{Spain}^{UK} = -13893.87$, $c_{Spain}^{UK} = 13748708.15$, and replacing them into equations (17) and (18). In this way, we can theoretically estimate the time at which the Spain reached a higher number of co-publications with UK than France as $t_{int} = 1998.9$. This methodology can be used with any set of $a_k^i, b_k^i, c_k^i$ coefficients, that represent the growth constants of the co-publications of country $k$ with a group of countries $i$. The last can be a very useful tool for S&T planners and for any other decision-making process.

## 7. Summary and discussion:

We have studied the long-term behavior of mainstream knowledge production for all Iberoamerican and Caribbean countries between 1973 and 2007. During the last decade, only two nations within the region (Spain and Brazil) were in the top-twenty most productive countries of the world. We showed that the production of scientific articles against time follows an exponential growth behavior, where Portugal, Colombia, Spain and Brazil, got the highest growth rates. The rest of the Iberoamerican nations presented a more irregular behavior that was directly correlated with their national macro-economical and political performances across the examined years.

We also analyzed the evolution of the national productivity, in terms of number of articles per million-inhabitants. Again, we found the excellent performance of the two European nations of our sample. Portugal increased its productivity 47 times during a 35-year period, and was seconded by Spain with 42 times. Both countries received the benefits of participating within the European Community Scientific Programs, which favored their continuous growth. Table 3 showed a very interesting difference between Spain and Portugal that might explain one of the causes of the higher growth slope of the last one. When we compare the traffic coefficient numbers (the ratio between the percentages of co-publications with larger scientific networks over the percentages of co-publications with smaller ones) Spain showed a 2.1 value, while Portugal the amazing 90.2 one. The strategy followed by Portugal to associate with hubs or other larger networks only (preferential attachment), got the prize of much more visibility, more opportunities to work in hot topics, great expansion of their connectivities and the possibility to be sharing with Mexico (ten times Portugal's population) the third rank of the region, in the number of mainstream publications per year.

Within Latin America, Chile, Argentina and Uruguay have the highest productivity, but their behavior presented some irregular shapes, that were in agreement with their internal economical and political performances during the same period. The productivity of the two most populated countries of the region, Brazil (188-million) and Mexico (107-million), are below the previously mentioned ones. But, they are also experiencing a continuous exponential growth-rate that will allow them to reach the highest Latin American productivities in just a few years. Another interesting case was Colombia, which after introducing some structural reforms to their national science and technology system, got the second nominal highest growth rate after Portugal. Even though, in the last case, the productivity still remains at almost the lowest place of our Iberoamerican sample (see Figure 3).

On the other hand, we have Jamaica and Venezuela, whose productivities remained practically constant during the whole 35-year period. We think that this is a good indicator of the absence of S&T policies or just their failure.

We also studied, with great detail, the co-publication profiles of 12 Iberoamerican countries that account for the 97.95 % of the whole regional mainstream knowledge production (1973-2006). Within the last sample we considered all the Iberoamerican countries in which their individual total production for the period was over 0.5% of the total aggregated number of publications for all Iberoamerican and Caribbean countries between 1973 and 2006 (see Figure 1). We analyzed the cooperation patterns of each country of the sample with other 45 different countries. We showed that USA, UK, France, Germany, Spain and Brazil, work like real hubs for all the Iberoamerican and Caribbean countries, concentrating the great majority of co-authorship activities.

We also found that most of the co-publications were performed with extra-regional countries. While, in the last decade, we detected a shift that favored the increase of intra-regional cooperation, as well as the emergence of new co-authorship networks with countries like South Korea, Russia, and China. The first might be explained by the existence of new regional cooperation agreements and programs, while the second by the globalization of knowledge production processes.

We showed that the scientific co-authorship among countries follows a power-law and behaves as a self-organizing scale-free network, where each country appears as a node and each co-publication as a link. We developed a mathematical model to study the temporal evolution of co-authorship networks, and we showed that the number of co-publications among countries growths quadratically against time. We determined the boundary conditions at which the self-organizing network process is triggered.

We empirically corroborated the quadratic growth prediction of our model by analyzing a 34-year temporal evolution of 352 different co-authorship networks. We showed how the number of co-publications $P_k^i(t)$ between country $k$ and country $i$, are related with a power-law against the coupling growth coefficients $a_k^i$ (scale-free). We calculated the quadratic growth coefficients $a_k^i, b_k^i, c_k^i$, for 352 different pairs of co-authorship countries across 34 years.

The last original results might be employed to estimate the future co-authorship behavior of mainstream scientific articles for the Iberoamerican countries. We have also presented a mathematical methodology, to use the empirically determined growth constants of each co-authorship network (see Appendix), to predict changes in the relative intensity of cooperation among different countries. All this constitutes a very useful set of tools that can be used by S&T planners and decision-makers worldwide.

From the data available at the Appendix, we calculated that the 47% of the networks have correlation coefficients with $R > 0.94$; 27 % with $0.94 > R > 0.90$; 18% with $0.90 > R > 0.84$ and 8 % with $R < 0.84$. These results show a very good agreement between our mathematical model and all the fitted data over a 34-year period.

We also corroborated the prediction that the connectivity of regional countries with larger scientific networks grows faster than with other less connected countries. We determined that 70.4 % of the 352 analyzed cases linked their cooperation with hubs or larger co-authorship networks. These social webs were responsible for 39.6 % of the total number of articles that were published, between 1973 and 2006, by the 12 countries of our sample. The rest of the co-authorship networks (smaller ones) accounted for only 8.8 % of the total regional production. Spain, alone, concentrated 6.1 % of the last 8.8%.

These smaller scientific co-authorship networks are composed by Latin American countries, where historical relationships, former colonial ties, same language and geographic proximity, can justify their existence. All these results are, again, consistent with the preferential attachment dynamics proposed as the bases of our mathematical model.

We think that the type of internal dynamics described by our mathematical model explains by itself the self-organizing property of the co-authorship scientific networks. Similar conclusions were obtained independently by Katz (1999) and Wagner & Leydesdorff (2005a). In order to estimate how the explicit S&T policy regional cooperation instruments work, or not, Lemarchand (2005) proposed to correlate output S&T indicators (i.e. bibliometric co-authorship) with the application of international cooperation agreements and programs. Unfortunately, the lack of regional databases with complete lists of all international scientific cooperation agreements among institutions and countries, did not allow us to test them in a systematic way.

In spite of the fact that we lacked complete regional databases, we used information provided by Lemarchand (2005, pp. 125-127), to analyze if there was any influence on intra-regional cooperation performance from those agreements between Argentina and the rest of Latin American & Caribbean countries (1965-2004). We found that the temporal evolution of formal S&T regional cooperative agreements (RCA) increases its size in a linear way, such as: RCA=3.5037$t$-6.7079 ($R$=0.997), where $t$ is time in years. The fact that RCA follows a linear growth like equation (1), and that the first might be embedded in the last, inhibit us to evaluate if the quadratic growth detected for the cooperation between Argentina and the rest of Iberoamerican countries, shown in Figure 9, was influenced by the regional cooperative agreements, or not. In order to formally reject its influence, we need to count each single mainstream scientific article that resulted from all these formal RCA. The last information was not available.

At this point we agree with Wagner and Leydesdorff (2005a) that the explicit S&T policies and other classical models can not account for the scientific co-authorship network dynamics. The analyses of 352 different scientific co-authorship networks over 34-year, provided in this article, present enough evidences to accept that the regional co-publication dynamics is a self-organizing process which is governed by preferential attachment.

We also applied our mathematical model to estimate the dates, $t_0$, at which the co-authorship connectivities trigger the self-organizing scale-free network for each of the 352 cases (see Appendix). We found that the last follows a normal distribution around year 1981.4 ± 2.2. We associated this particular concentration of dates which trigger the self-organizing co-authorship networks, with a massive brain-drainage caused by the adverse regional political situation, between the mid-sixties and mid-eighties, where more than 70% of the Iberoamerican and Caribbean population were living under Dictatorship's Governments (see Figure 18). We have examined, both, the pulling and pushing factors for the international mobility and migration of talents, during those days.

The examined data showed a time-lag of ≈15 years, between the peak when most of the Iberoamerican population was living under dictatorships (massive brain drainage) and the peak when most of the co-authorship networks were triggered. We think that emigrant scientists while living abroad need some period of time to develop a wide variety of S&T human capital assets, not only enhanced S&T knowledge, but craft skills, know-how, publish mainstream articles, develop their ability to structure and plan research and of course, increase contacts with other scientists, industry, and funding agents. After expanding these potentialities, the emigrant scientist became "visible" for their home-country scientific networks. After this time-lag, they might be in a position to start transferring part of their accumulated knowledge and experience to their home country through periodic visits and by their participation in knowledge and co-authorship networks.

In recent years, several Latin American countries, established all sort of national programs (i.e. CALDAS in Colombia; RAICES in Argentina; Red de Talentos para la Innovación in Mexico, Chile Global; etc.) with the strategic goal to coordinate and stimulate the scientific Diasporas in order to strengthen knowledge networks with their home countries. These S&T policies have also the possibility to enhance the self-organizing mechanisms of co-authorship networks and transform brain-drain into brain-gain.

# Appendix: Table 4 to 15

## Table 4: Main co-authorship networks of Spain

| No. | Country | Co-publications (1973-2006) | % | α | B | C | $T_o$ | R |
|-----|---------|------------------|------|--------|------------|----------------|-------|-------|
| 1 | USA | 37137 | 7.44 | 5.2616 | -20824.690 | 20605240.8702 | 1979 | 0.997 |
| 2 | France | 24697 | 4.95 | 2.8737 | -11364.005 | 11234641.9734 | 1977 | 0.995 |
| 3 | United Kingdom | 24009 | 4.81 | 3.5101 | -13893.870 | 13748788.1501 | 1979 | 0.995 |
| 4 | Germany | 18944 | 3.80 | 3.3702 | -13349.061 | 13218467.0496 | 1980 | 0.992 |
| 5 | Italy | 17122 | 3.43 | 3.0490 | -12076.616 | 11958433.6323 | 1980 | 0.992 |
| 6 | Netherlands | 8615 | 1.73 | 1.5474 | -6128.9265 | 6068996.4219 | 1980 | 0.989 |
| 7 | Switzerland | 6612 | 1.33 | 0.9903 | -3920.2239 | 3879575.4657 | 1979 | 0.995 |
| 8 | Belgium | 6125 | 1.23 | 1.0162 | -4024.2917 | 3984234.4165 | 1980 | 0.990 |
| 9 | Canada | 5823 | 1.17 | 1.2080 | -4787.2990 | 4743055.6354 | 1981 | 0.969 |
| 10 | Sweden | 4982 | 1.00 | 0.8608 | -3409.2103 | 3375406.6912 | 1980 | 0.991 |
| 11 | Portugal | 4589 | 0.92 | 1.1506 | -4561.8716 | 4521508.4837 | 1982 | 0.976 |
| 12 | Mexico | 4472 | 0.90 | 0.9578 | -3795.7427 | 3760779.8047 | 1981 | 0.985 |
| 13 | Argentina | 4301 | 0.86 | 0.8276 | -3278.6325 | 3247044.6473 | 1981 | 0.994 |
| 14 | URSS/Russia | 4241 | 0.85 | 0.7547 | -2988.4882 | 2958514.6452 | 1980 | 0.973 |
| 15 | Denmark | 3820 | 0.77 | 0.5922 | -2344.0977 | 2319713.9950 | 1979 | 0.985 |
| 16 | Japan | 3303 | 0.66 | 0.6662 | -2639.7982 | 2614872.2964 | 1981 | 0.988 |
| 17 | Brazil | 3259 | 0.65 | 0.6344 | -2513.4417 | 2489352.0585 | 1981 | 0.987 |
| 18 | Poland | 3160 | 0.63 | 0.4687 | -1858.0196 | 1841212.2366 | 1982 | 0.960 |
| 19 | Austria | 3091 | 0.62 | 0.6009 | -2380.5616 | 2357847.0327 | 1981 | 0.986 |
| 20 | Finland | 2685 | 0.54 | 0.4689 | -1856.9452 | 1838306.2111 | 1980 | 0.983 |
| 21 | Chile | 2643 | 0.53 | 0.5025 | -1986.9815 | 1968246.9592 | 1977 | 0.970 |
| 22 | Australia | 2338 | 0.47 | 0.6033 | -2392.1201 | 2371271.5908 | 1983 | 0.967 |
| 23 | Norway | 2040 | 0.41 | 0.4687 | -1858.0196 | 1841212.2366 | 1982 | 0.960 |
| 24 | Israel | 2037 | 0.41 | 0.3857 | -1527.9671 | 1513194.8183 | 1981 | 0.981 |
| 25 | China | 1918 | 0.38 | 0.3860 | -1529.4483 | 1514941.1227 | 1981 | 0.987 |
| 26 | Ireland | 1557 | 0.31 | 0.3825 | -1516.4231 | 1503009.1234 | 1982 | 0.974 |
| 27 | India | 1331 | 0.27 | 0.2241 | -887.5645 | 878761.0904 | 1980 | 0.977 |
| 28 | Cuba | 1202 | 0.24 | 0.2641 | -1046.6025 | 1036850.7810 | 1981 | 0.986 |
| 29 | South Korea | 1087 | 0.22 | 0.2096 | -830.1223 | 822020.2571 | 1980 | 0.967 |
| 30 | Venezuela | 1033 | 0.21 | 0.2126 | -842.7931 | 835068.8973 | 1982 | 0.973 |
| 31 | Colombia | 996 | 0.20 | 0.2818 | -1117.5716 | 1108019.0301 | 1983 | 0.964 |
| 32 | Peru | 449 | 0.09 | 0.0910 | -360.5324 | 357140.4285 | 1981 | 0.961 |
| 33 | Uruguay | 418 | 0.08 | 0.0966 | -383.0340 | 379542.6738 | 1983 | 0.976 |
| 34 | Costa Rica | 176 | 0.04 | 0.0331 | -131.1052 | 139837.3730 | 1980 | 0.948 |
| 35 | Panamá | 132 | 0.03 | 0.0319 | -126.6023 | 125427.1437 | 1984 | 0.959 |
| 36 | Bolivia | 125 | 0.03 | 0.0260 | -103.0420 | 102069.2045 | 1982 | 0.918 |
| 37 | Ecuador | 84 | 0.02 | 0.0156 | -61.7276 | 61127.8803 | 1978 | 0.852 |
| 38 | Paraguay | 66 | 0.01 | 0.0141 | -55.8192 | 55333.7371 | 1979 | 0.762 |
| 39 | Guatemala | 40 | 0.01 | 0.0164 | -65.1970 | 64898.0900 | 1988 | 0.749 |

Table 5: Main co-authorship networks of Brazil

| No. | Country | Co-publications (1973-2006) | % | $a$ | $b$ | $c$ | $T_o$ | $R$ |
|---|---|---|---|---|---|---|---|---|
| 1 | USA | 26662 | 11.60 | 0.9775 | -3866.117 | 3822666.5106 | 1978 | 0.997 |
| 2 | France | 8511 | 3.70 | 0.9897 | -3915.519 | 3872902.1899 | 1978 | 0.992 |
| 3 | United Kingdom | 8031 | 3.49 | 1.0436 | -4132.414 | 4091037.4820 | 1980 | 0.989 |
| 4 | Germany | 6723 | 2.93 | 0.5826 | -2305.617 | 2281066.0570 | 1979 | 0.994 |
| 5 | Italy | 4344 | 1.89 | 0.6344 | -2513.442 | 2489352.0585 | 1981 | 0.987 |
| 6 | Canada | 4155 | 1.81 | 0.5618 | -2224.166 | 2201380.1368 | 1979 | 0.986 |
| 7 | Argentina | 3338 | 1.45 | 0.6522 | -2585.684 | 2562686.0997 | 1982 | 0.981 |
| 8 | Spain | 3259 | 1.42 | 0.6344 | -2513.442 | 2489352.0585 | 1981 | 0.987 |
| 9 | Japan | 2296 | 1.00 | 0.3568 | -1412.698 | 1398403.1486 | 1980 | 0.994 |
| 10 | URSS/Russia | 1981 | 0.86 | 0.2985 | -1181.060 | 1168400.3928 | 1978 | 0.954 |
| 11 | Netherlands | 1858 | 0.81 | 0.3600 | -1426.427 | 1412946.5958 | 1981 | 0.983 |
| 12 | Portugal | 1839 | 0.80 | 0.4384 | -1737.931 | 1722424.4280 | 1982 | 0.974 |
| 13 | Mexico | 1659 | 0.72 | 0.2899 | -148.6073 | 1137624.2447 | 1981 | 0.977 |
| 14 | Belgium | 1635 | 0.71 | 0.2056 | -813.2625 | 804179.9963 | 1978 | 0.988 |
| 15 | Switzerland | 1626 | 0.71 | 0.1913 | -756.4893 | 747989.4049 | 1977 | 0.970 |
| 16 | Sweden | 1452 | 0.63 | 0.1914 | -757.5026 | 749305.2241 | 1979 | 0.985 |
| 17 | Chile | 1440 | 0.63 | 0.2031 | -804.0688 | 795742.3052 | 1979 | 0.970 |
| 18 | Australia | 1373 | 0.60 | 0.3314 | -1313.979 | 1302358.6141 | 1982 | 0.972 |
| 19 | India | 1086 | 0.47 | 0.2249 | -891.4852 | 883404.2768 | 1982 | 0.963 |
| 20 | China | 1081 | 0.47 | 0.2749 | -1089.808 | 1080158.0707 | 1982 | 0.984 |
| 21 | Poland | 1057 | 0.46 | 0.1192 | -471.2907 | 465724.4286 | 1977 | 0.967 |
| 22 | Israel | 998 | 0.43 | 0.1645 | -651.5691 | 645318.5184 | 1980 | 0.947 |
| 23 | Denmark | 831 | 0.36 | 0.0859 | -339.3886 | 335407.5107 | 1975 | 0.950 |
| 24 | Colombia | 810 | 0.35 | 0.1571 | -622.3704 | 616581.0912 | 1981 | 0.969 |
| 25 | Austria | 716 | 0.31 | 0.0908 | -358.0885 | 355063.2028 | 1972 | 0.952 |
| 26 | Finland | 614 | 0.27 | 0.0538 | -212.3350 | 209498.3882 | 1973 | 0.948 |
| 27 | Norway | 609 | 0.26 | 0.0559 | -220.6054 | 217699.6396 | 1973 | 0.928 |
| 28 | Venezuela | 604 | 0.26 | 0.0888 | -351.5691 | 348100.5231 | 1980 | 0.954 |
| 29 | Uruguay | 492 | 0.21 | 0.1201 | -476.3842 | 472294.7166 | 1983 | 0.969 |
| 30 | South Korea | 472 | 0.21 | 0.1351 | -535.9280 | 531337.9024 | 1983 | 0.954 |
| 31 | Jamaica | 470 | 0.20 | 0.1145 | -453.9120 | 448842.3347 | 1982 | 0.964 |
| 32 | Cuba | 469 | 0.20 | 0.1132 | -448.8316 | 444774.5477 | 1982 | 0.967 |
| 33 | Peru | 341 | 0.15 | 0.0716 | -283.9686 | 281449.0659 | 1983 | 0.897 |
| 34 | Ireland | 301 | 0.13 | 0.0806 | -319.7258 | 316993.6786 | 1983 | 0.871 |
| 35 | Costa Rica | 205 | 0.09 | 0.0471 | -186.9644 | 185375.6686 | 1985 | 0.895 |
| 36 | Ecuador | 197 | 0.09 | 0.0650 | -257.9493 | 255854.6161 | 1984 | 0.891 |
| 37 | Panamá | 120 | 0.05 | 0.0396 | -157.2140 | 155950.6608 | 1985 | 0.896 |
| 38 | Bolivia | 106 | 0.05 | 0.0171 | -67.7170 | 67036.2021 | 1980 | 0.866 |
| 39 | Paraguay | 35 | 0.02 | 0.0185 | -73.4062 | 72747.7876 | 1984 | 0.815 |
| 40 | Guatemala | 32 | 0.01 | 0.0070 | -27.7442 | 27510.9331 | 1982 | 0.758 |

Table 6: Main co-authorship networks of Mexico

| No. | Country | Co-publications (1973-2006) | % | $a$ | $b$ | $c$ | $T_o$ | $R$ |
|---|---|---|---|---|---|---|---|---|
| 1 | USA | 73930 | 46.28 | 2.3376 | -9208.255 | 9063205.9000 | 1970 | 0.988 |
| 2 | Spain | 4472 | 2.80 | 0.9578 | -795.7427 | 3760779.8047 | 1981 | 0.985 |
| 3 | France | 4337 | 2.71 | 0.6318 | -2501.203 | 2475493.1531 | 1979 | 0.990 |
| 4 | Canada | 3638 | 2.28 | 0.5076 | -2009.722 | 1989358.3726 | 1980 | 0.977 |
| 5 | Germany | 3594 | 2.25 | 0.5995 | -2374.226 | 2350670.5611 | 1980 | 0.990 |
| 6 | United Kingdom | 3187 | 1.99 | 0.5123 | -2028.922 | 2008770.4382 | 1980 | 0.978 |
| 7 | Italy | 2173 | 1.36 | 0.3789 | -1500.861 | 1486202.1795 | 1981 | 0.989 |
| 8 | URSS/Russia | 1996 | 1.25 | 0.4149 | -1643.838 | 1628246.5704 | 1981 | 0.973 |
| 9 | Brazil | 1659 | 1.04 | 0.2899 | -1148.607 | 1137624.2447 | 1981 | 0.977 |
| 10 | Japan | 1625 | 1.02 | 0.2698 | -068.7510 | 1058219.4341 | 1981 | 0.975 |
| 11 | Argentina | 1162 | 0.73 | 0.2275 | -901.6184 | 893140.7198 | 1982 | 0.972 |
| 12 | Australia | 1082 | 0.68 | 0.2094 | -829.9640 | 822303.4951 | 1982 | 0.979 |
| 13 | Switzerland | 1051 | 0.66 | 0.1931 | -765.4875 | 758503.7546 | 1982 | 0.961 |
| 14 | Netherlands | 1016 | 0.64 | 0.1962 | -777.5430 | 770234.2759 | 1982 | 0.973 |
| 15 | Poland | 903 | 0.57 | 0.1166 | -461.1519 | 456037.5995 | 1977 | 0.961 |
| 16 | China | 893 | 0.56 | 0.2519 | -999.1780 | 990652.7378 | 1983 | 0.962 |
| 17 | Cuba | 880 | 0.55 | 0.1589 | -629.4325 | 623134.0570 | 1981 | 0.928 |
| 18 | India | 849 | 0.53 | 0.1488 | -589.6196 | 583928.3120 | 1981 | 0.962 |
| 19 | Sweden | 835 | 0.52 | 0.1757 | -696.7213 | 690573.8430 | 1983 | 0.942 |
| 20 | Chile | 786 | 0.49 | 0.2899 | -1148.607 | 1137624.2447 | 1981 | 0.977 |
| 21 | South Korea | 757 | 0.47 | 0.2374 | -941.7514 | 933915.3567 | 1983 | 0.945 |
| 22 | Israel | 713 | 0.45 | 0.0803 | -317.8334 | 314343.4214 | 1979 | 0.937 |
| 23 | Belgium | 707 | 0.44 | 0.1429 | -563.0165 | 557802.4722 | 1970 | 0.979 |
| 24 | Colombia | 684 | 0.43 | 0.1019 | -403.1278 | 398895.7063 | 1978 | 0.947 |
| 25 | Venezuela | 518 | 0.32 | 0.1242 | -492.4340 | 488212.6706 | 1982 | 0.948 |
| 26 | Denmark | 399 | 0.25 | 0.0313 | -123.4911 | 121845.7708 | 1973 | 0.949 |
| 27 | Costa Rica | 305 | 0.19 | 0.0437 | -173.0405 | 171460.1209 | 1980 | 0.878 |
| 28 | Finland | 303 | 0.19 | 0.0837 | -331.9970 | 329185.9228 | 1983 | 0.866 |
| 29 | Ireland | 270 | 0.17 | 0.0656 | -260.1802 | 257901.8885 | 1983 | 0.871 |
| 30 | Austria | 259 | 0.16 | 0.0419 | -165.8507 | 164191.8885 | 1979 | 0.963 |
| 31 | Ecuador | 217 | 0.14 | 0.0686 | -271.9997 | 269791.6400 | 1983 | 0.876 |
| 32 | Norway | 195 | 0.12 | 0.0325 | -128.8304 | 127643.0443 | 1982 | 0.921 |
| 33 | Portugal | 191 | 0.12 | 0.0595 | -236.1744 | 234204.0948 | 1985 | 0.942 |
| 34 | Peru | 163 | 0.10 | 0.0416 | -164.9562 | 163543.3600 | 1983 | 0.909 |
| 35 | Uruguay | 156 | 0.10 | 0.0315 | -124.7898 | 123634.9285 | 1981 | 0.950 |
| 36 | Guatemala | 147 | 0.09 | 0.0180 | -71.0261 | 70250.4119 | 1973 | 0.882 |
| 37 | Bolivia | 66 | 0.04 | 0.0141 | -56.0152 | 55503.1473 | 1986 | 0.913 |
| 38 | Honduras | 59 | 0.04 | 0.0131 | -52.0809 | 51612.9570 | 1988 | 0.828 |
| 39 | Jamaica | 44 | 0.03 | 0.0030 | -11.8538 | 11709.6191 | 1976 | 0.651 |
| 40 | Panamá | 35 | 0.02 | 0.0160 | -41.9125 | 41556.1178 | 1981 | 0.821 |

Table 7: Main co-authorship networks of Argentina

| No. | Country | Co-publications (1973-2006) | % | $a$ | $b$ | $c$ | $T_o$ | $R$ |
|---|---|---|---|---|---|---|---|---|
| 1 | USA | 10178 | 9.77 | 1.3437 | -5320.097 | 5265924.9431 | 1980 | 0.990 |
| 2 | Spain | 4301 | 4.13 | 0.8276 | -3278.633 | 3247044.6433 | 1981 | 0.994 |
| 3 | Brazil | 3338 | 3.20 | 0.6522 | -2585.684 | 2562686.0097 | 1982 | 0.979 |
| 4 | France | 3118 | 2.99 | 0.3644 | -441.3602 | 1425463.7434 | 1978 | 0.992 |
| 5 | Germany | 2773 | 2.66 | 0.3877 | -1534.665 | 1518636.2753 | 1979 | 0.986 |
| 6 | United Kingdom | 2197 | 2.11 | 0.3998 | -1584.381 | 1569587.4631 | 1981 | 0.975 |
| 7 | Italy | 1854 | 1.78 | 0.3120 | -236.4142 | 1225028.4042 | 1981 | 0.995 |
| 8 | Canada | 1348 | 1.29 | 0.2663 | -1055.265 | 1045492.6103 | 1981 | 0.969 |
| 9 | Chile | 1309 | 1.26 | 0.2725 | -1080.515 | 1070936.9381 | 1983 | 0.962 |
| 10 | Mexico | 1162 | 1.11 | 0.2275 | -901.6184 | 893140.7198 | 1982 | 0.972 |
| 11 | Netherlands | 731 | 0.70 | 0.1480 | -586.5066 | 581101.8342 | 1981 | 0.961 |
| 12 | Sweden | 668 | 0.64 | 0.0479 | -188.9380 | 186260.4336 | 1972 | 0.968 |
| 13 | Uruguay | 625 | 0.60 | 0.1359 | -538.7078 | 533849.0381 | 1982 | 0.974 |
| 14 | Australia | 596 | 0.57 | 0.1300 | -515.3506 | 510656.2496 | 1982 | 0.970 |
| 15 | Switzerland | 593 | 0.57 | 0.1032 | -408.8752 | 404941.3388 | 1981 | 0.913 |
| 16 | Japan | 592 | 0.57 | 0.0956 | -378.5930 | 374745.3267 | 1980 | 0.982 |
| 17 | Belgium | 516 | 0.50 | 0.0922 | -365.5657 | 362179.3842 | 1982 | 0.970 |
| 18 | Colombia | 473 | 0.45 | 0.1224 | -485.2650 | 481067.2324 | 1982 | 0.924 |
| 19 | Venezuela | 390 | 0.37 | 0.0485 | -192.0244 | 190111.6180 | 1980 | 0.953 |
| 20 | URSS/Russia | 362 | 0.35 | 0.0762 | -302.0084 | 299165.2511 | 1982 | 0.933 |
| 21 | Israel | 356 | 0.34 | 0.0455 | -180.1836 | 178283.7788 | 1980 | 0.921 |
| 22 | Denmark | 345 | 0.33 | 0.0468 | -185.1462 | 183305.7071 | 1978 | 0.952 |
| 23 | Austria | 331 | 0.32 | 0.0743 | -294.3177 | 291643.9181 | 1981 | 0.972 |
| 24 | India | 331 | 0.32 | 0.0820 | -325.0428 | 322186.5795 | 1982 | 0.924 |
| 25 | China | 291 | 0.28 | 0.0896 | -355.3495 | 352368.0823 | 1983 | 0.916 |
| 26 | Poland | 288 | 0.28 | 0.0557 | -220.4897 | 218341.8293 | 1979 | 0.933 |
| 27 | Finland | 237 | 0.23 | 0.0059 | -230.5183 | 22794.9867 | 1986 | 0.880 |
| 28 | South Korea | 226 | 0.22 | 0.0578 | -229.0681 | 227028.2068 | 1982 | 0.845 |
| 29 | Portugal | 222 | 0.21 | 0.0847 | -336.0897 | 333420.1699 | 1984 | 0.898 |
| 30 | Peru | 190 | 0.18 | 0.0537 | -213.1090 | 211340.7053 | 1984 | 0.907 |
| 31 | Ireland | 172 | 0.17 | 0.0637 | -252.9253 | 250929.7146 | 1985 | 0.864 |
| 32 | Ecuador | 161 | 0.15 | 0.0583 | -231.2614 | 229405.4537 | 1983 | 0.866 |
| 33 | Cuba | 149 | 0.14 | 0.0406 | -161.2350 | 159887.8445 | 1986 | 0.929 |
| 34 | Bolivia | 100 | 0.10 | 0.0252 | -100.0841 | 99239.1310 | 1986 | 0.903 |
| 35 | Norway | 89 | 0.09 | 0.0207 | -82.1210 | 81392.098 | 1984 | 0.897 |
| 36 | Costa Rica | 85 | 0.08 | 0.0222 | -88.0110 | 87251.5958 | 1982 | 0.842 |
| 37 | Paraguay | 77 | 0.07 | 0.0177 | -70.0916 | 69495.5021 | 1980 | 0.864 |
| 38 | Guatemala | 32 | 0.03 | 0.0052 | -20.4378 | 20250.9787 | 1965 | 0.719 |
| 39 | Panamá | 27 | 0.03 | 0.0084 | -33.4097 | 33144.1539 | 1989 | 0.749 |
| 40 | Honduras | 11 | 0.01 | 0.0033 | -13.2580 | 13147.8193 | 2009 | 0.688 |

Table 8: Main co-authorship networks of Portugal

| No. | Country | Co-publications (1973-2006) | % | *a* | *b* | *c* | *T$_o$* | *R* |
|---|---|---|---|---|---|---|---|---|
| 1 | United Kingdom | 6901 | 10.86 | 1.0252 | -4058.957 | 4017347.1004 | 1980 | 0.991 |
| 2 | USA | 6079 | 9.56 | 1.0483 | -152.603 | 4112537.7950 | 1981 | 0.988 |
| 3 | France | 5186 | 8.16 | 0.7807 | -3090.709 | 3059067.7837 | 1979 | 0.995 |
| 4 | Spain | 4589 | 7.22 | 1.1506 | -4561.872 | 4521508.4837 | 1982 | 0.976 |
| 5 | Germany | 3677 | 5.78 | 0.7991 | -3166.786 | 3137617.5876 | 1981 | 0.986 |
| 6 | Italy | 2673 | 4.20 | 0.5292 | -2096.416 | 2076384.3826 | 1981 | 0.990 |
| 7 | Netherlands | 2234 | 3.51 | 0.3372 | -1334.419 | 1320338.6973 | 1979 | 0.988 |
| 8 | Brazil | 1839 | 2.89 | 0.4384 | -1737.931 | 1722424.4280 | 1982 | 0.974 |
| 9 | Belgium | 1597 | 2.51 | 0.2278 | -901.5372 | 891889.4761 | 1979 | 0.981 |
| 10 | Sweden | 1564 | 2.46 | 0.2565 | -1015.403 | 1001999.3458 | 1979 | 0.983 |
| 11 | Switzerland | 1533 | 2.41 | 0.2644 | -1047.022 | 1036625.8457 | 1980 | 0.978 |
| 12 | URSS/Russia | 1198 | 1.88 | 0.2827 | -120.3647 | 1110148.6091 | 1982 | 0.987 |
| 13 | Denmark | 930 | 1.46 | 0.1153 | -455.9256 | 450648.9168 | 1977 | 0.934 |
| 14 | Austria | 901 | 1.42 | 0.2025 | -802.4929 | 795111.0959 | 1981 | 0.973 |
| 15 | Canada | 896 | 1.41 | 0.1644 | -651.3962 | 645160.1825 | 1981 | 0.972 |
| 16 | Finland | 854 | 1.34 | 0.1561 | -618.0995 | 612020.6406 | 1980 | 0.981 |
| 17 | Poland | 827 | 1.30 | 0.1734 | -686.9567 | 680533.4122 | 1981 | 0.958 |
| 18 | Norway | 783 | 1.23 | 0.0686 | -270.9538 | 267412.4284 | 1975 | 0.960 |
| 19 | Japan | 598 | 0.94 | 0.1433 | -568.0787 | 563019.7082 | 1982 | 0.957 |
| 20 | China | 480 | 0.76 | 0.1612 | -639.5416 | 634288.8148 | 1984 | 0.953 |
| 21 | Ireland | 472 | 0.74 | 0.0953 | -377.7270 | 374183.1447 | 1982 | 0.982 |
| 22 | Israel | 392 | 0.62 | 0.0908 | -359.9498 | 356735.7409 | 1982 | 0.938 |
| 23 | Australia | 341 | 0.54 | 0.1004 | -398.2632 | 394940.2546 | 1983 | 0.945 |
| 24 | Argentina | 222 | 0.35 | 0.0847 | -336.0897 | 333420.1699 | 1984 | 0.898 |
| 25 | India | 217 | 0.34 | 0.0748 | -296.6035 | 294203.6272 | 1983 | 0.930 |
| 26 | Mexico | 191 | 0.30 | 0.0595 | -236.1744 | 234204.0948 | 1985 | 0.942 |
| 27 | Chile | 171 | 0.27 | 0.0642 | -254.7320 | 252697.2250 | 1984 | 0.894 |
| 28 | South Korea | 68 | 0.11 | 0.0243 | -96.5520 | 95770.2431 | 1987 | 0.885 |
| 29 | Venezuela | 48 | 0.08 | 0.0115 | -45.5740 | 45187.3311 | 1981 | 0.806 |
| 30 | Cuba | 44 | 0.07 | 0.0026 | -10.5158 | 10432.4057 | 1996 | 0.677 |
| 31 | Colombia | 39 | 0.06 | 0.0125 | -49.5653 | 49159.0070 | 1983 | 0.823 |

Table 9: Main co-authorship networks of Chile

| No. | Country | Co-publications (1973-2006) | % | $a$ | $b$ | $c$ | $T_o$ | $R$ |
|---|---|---|---|---|---|---|---|---|
| 1 | USA | 8165 | 14.95 | 0.9264 | -3667.919 | 3630717.4683 | 1980 | 0.980 |
| 2 | Spain | 2643 | 4.84 | 0.5015 | -1986.982 | 1968246.9592 | 1981 | 0.970 |
| 3 | Germany | 2542 | 4.65 | 0.3975 | -574.2888 | 1558812.0291 | 1980 | 0.984 |
| 4 | France | 2500 | 4.58 | 0.4311 | -1707.676 | 1691261.6081 | 1981 | 0.986 |
| 5 | United Kingdom | 1815 | 3.32 | 0.3034 | -1202.252 | 1190870.8707 | 1981 | 0.978 |
| 6 | Brazil | 1440 | 2.64 | 0.2021 | -804.0688 | 795742.3052 | 1989 | 0.970 |
| 7 | Argentina | 1309 | 2.40 | 0.2725 | -1080.515 | 1070936.9381 | 1983 | 0.962 |
| 8 | Canada | 1107 | 2.03 | 0.1322 | -523.4935 | 518177.9914 | 1980 | 0.925 |
| 9 | Italy | 1085 | 1.99 | 0.2052 | -813.2239 | 805630.4123 | 1982 | 0.966 |
| 10 | Mexico | 786 | 1.44 | 0.1431 | -566.9429 | 501521.5420 | 1981 | 0.975 |
| 11 | Australia | 600 | 1.10 | 0.1261 | -499.9402 | 495526.0404 | 1982 | 0.953 |
| 12 | Belgium | 565 | 1.03 | 0.0847 | -335.5056 | 332362.7517 | 1981 | 0.941 |
| 13 | Netherlands | 562 | 1.03 | 0.1150 | -455.7427 | 451660.3041 | 1981 | 0.948 |
| 14 | Sweden | 510 | 0.93 | 0.0591 | -233.5962 | 231005.4567 | 1976 | 0.976 |
| 15 | URSS/Russia | 464 | 0.85 | 0.1551 | -615.3990 | 610321.4515 | 1984 | 0.931 |
| 16 | Switzerland | 452 | 0.83 | 0.0851 | -337.0815 | 333982.8325 | 1981 | 0.900 |
| 17 | Japan | 400 | 0.73 | 0.0685 | -271.4356 | 268800.7619 | 1981 | 0.966 |
| 18 | Denmark | 299 | 0.55 | 0.0733 | -290.8086 | 288346.9360 | 1984 | 0.954 |
| 19 | Israel | 298 | 0.55 | 0.0527 | -208.9247 | 206893.4434 | 1982 | 0.927 |
| 20 | Venezuela | 250 | 0.46 | 0.0267 | -104.8874 | 103728.4688 | 1964 | 0.920 |
| 21 | Poland | 248 | 0.45 | 0.0663 | -262.7491 | 206491.4833 | 1982 | 0.928 |
| 22 | Colombia | 240 | 0.44 | 0.0494 | -195.7159 | 193905.2759 | 1981 | 0.950 |
| 23 | Finland | 216 | 0.40 | 0.0462 | -182.9613 | 181317.4907 | 1980 | 0.927 |
| 24 | China | 214 | 0.39 | 0.0449 | -177.9904 | 176310.1910 | 1982 | 0.965 |
| 25 | Uruguay | 209 | 0.38 | 0.0456 | -180.5321 | 178886.6309 | 1980 | 0.961 |
| 26 | Austria | 208 | 0.38 | 0.0425 | -168.2940 | 166751.9833 | 1980 | 0.927 |
| 27 | India | 199 | 0.36 | 0.0353 | -139.9936 | 138638.9867 | 1983 | 0.920 |
| 28 | Ecuador | 177 | 0.32 | 0.0626 | -248.2483 | 246286.8774 | 1983 | 0.890 |
| 29 | Portugal | 171 | 0.31 | 0.0642 | -254.7320 | 252697.2250 | 1984 | 0.894 |
| 30 | Peru | 170 | 0.31 | 0.0373 | -148.0655 | 146822.6899 | 1985 | 0.872 |
| 31 | South Korea | 136 | 0.25 | 0.0482 | -191.2781 | 189734.9498 | 1984 | 0.925 |
| 32 | Cuba | 112 | 0.21 | 0.0269 | -106.6625 | 105743.8905 | 1983 | 0.936 |
| 33 | Ireland | 96 | 0.18 | 0.0286 | -113.4285 | 112495.4026 | 1983 | 0.784 |
| 34 | Norway | 84 | 0.15 | 0.0134 | -53.2756 | 52762.1228 | 1988 | 0.831 |
| 35 | Bolivia | 56 | 0.10 | 0.0107 | -42.3591 | 42035.8036 | 1979 | 0.718 |

Table 10: Main co-authorship networks of Venezuela

| No. | Country | Co-publications (1973-2006) | % | $a$ | $b$ | $c$ | $T_o$ | $R$ |
|-----|---------|-----|-----|-----|-----|-----|-----|-----|
| 1 | USA | 4287 | 16.17 | 0.2238 | -883.8616 | 872583.5826 | 1975 | 0.959 |
| 2 | France | 1128 | 4.26 | 0.0832 | -328.1988 | 323638.1017 | 1972 | 0.971 |
| 3 | Spain | 1033 | 3.90 | 0.2126 | -842.7931 | 835068.8973 | 1982 | 0.973 |
| 4 | United Kingdom | 984 | 3.71 | 0.0685 | -270.6737 | 267428.0335 | 1976 | 0.953 |
| 5 | Brazil | 604 | 2.28 | 0.0888 | -351.5691 | 348100.5231 | 1980 | 0.954 |
| 6 | Mexico | 518 | 1.95 | 0.1242 | -492.4340 | 488212.6706 | 1982 | 0.948 |
| 7 | Italy | 492 | 1.86 | 0.0295 | -116.0418 | 114304.3373 | 1967 | 0.935 |
| 8 | Germany | 453 | 1.71 | 0.0629 | -248.9092 | 246412.7712 | 1979 | 0.953 |
| 9 | Canada | 430 | 1.62 | 0.0238 | -93.7329 | 92377.2466 | 1969 | 0.904 |
| 10 | Argentina | 390 | 1.47 | 0.0401 | -158.8303 | 157126.8055 | 1980 | 0.940 |
| 11 | Chile | 250 | 0.94 | 0.0265 | -104.8874 | 103728.4688 | 1979 | 0.920 |
| 12 | Colombia | 222 | 0.84 | 0.0470 | -186.1174 | 184409.4761 | 1980 | 0.967 |
| 13 | Japan | 171 | 0.65 | 0.0037 | -14.3695 | 14072.4396 | 1942 | 0.644 |
| 14 | Sweden | 139 | 0.52 | 0.0089 | -35.3582 | 35033.3542 | 1986 | 0.748 |
| 15 | Switzerland | 134 | 0.51 | 0.0139 | -55.0392 | 54382.2580 | 1980 | 0.851 |
| 16 | Belgium | 130 | 0.49 | 0.0328 | -130.0404 | 129009.1222 | 1982 | 0.915 |
| 17 | Norway | 124 | 0.47 | 0.0196 | -77.7400 | 76960.7969 | 1983 | 0.951 |
| 18 | Netherlands | 120 | 0.45 | 0.0222 | -87.8018 | 86994.0361 | 1978 | 0.882 |
| 19 | Australia | 113 | 0.43 | 0.0184 | -73.0062 | 72299.1095 | 1984 | 0.930 |
| 20 | Poland | 110 | 0.42 | 0.0048 | -19.4450 | 19531.6489 | 1996 | 0.736 |
| 21 | Israel | 93 | 0.35 | 0.0073 | -29.0204 | 28728.8986 | 1988 | 0.660 |
| 22 | URSS/Russia | 89 | 0.34 | 0.0145 | -57.5772 | 56973.5606 | 1985 | 0.847 |
| 23 | Cuba | 82 | 0.31 | 0.0134 | -52.8952 | 52351.7649 | 1974 | 0.870 |
| 24 | Peru | 75 | 0.28 | 0.0147 | -58.1914 | 57691.0943 | 1979 | 0.711 |
| 25 | Uruguay | 67 | 0.25 | 0.0120 | -47.3477 | 46889.7717 | 1973 | 0.829 |

Table 11: Main co-authorship networks of Colombia

| No. | Country | Co-publications (1973-2006) | % | $a$ | $b$ | $c$ | $T_o$ | $R$ |
|-----|---------|------------|-------|--------|-----------|---------------|------|-------|
| 1 | USA | 3347 | 26.62 | 0.3492 | -1381.448 | 1366468.3235 | 1978 | 0.978 |
| 2 | Spain | 996 | 7.92 | 0.2818 | -117.5716 | 1108019.0301 | 1983 | 0.964 |
| 3 | United Kingdom | 834 | 6.63 | 0.2078 | -824.1178 | 816901.8382 | 1983 | 0.950 |
| 4 | Brazil | 810 | 6.44 | 0.1571 | -622.3704 | 616581.0912 | 1981 | 0.969 |
| 5 | France | 699 | 5.56 | 0.1245 | -493.1534 | 488294.7602 | 1981 | 0.957 |
| 6 | Mexico | 684 | 5.44 | 0.1019 | -403.1278 | 398895.7063 | 1978 | 0.947 |
| 7 | Germany | 500 | 3.98 | 0.1140 | -452.1409 | 448209.3539 | 1983 | 0.935 |
| 8 | Argentina | 473 | 3.76 | 0.1224 | -485.2650 | 481067.2324 | 1982 | 0.924 |
| 9 | Canada | 333 | 2.65 | 0.0722 | -286.1228 | 283561.3152 | 1981 | 0.856 |
| 10 | Netherlands | 266 | 2.12 | 0.0747 | -296.0811 | 293575.2299 | 1982 | 0.898 |
| 11 | India | 263 | 2.09 | 0.0587 | -232.5191 | 230398.6834 | 1981 | 0.912 |
| 12 | URSS/Russia | 261 | 2.08 | 0.0429 | -169.7758 | 168000.5384 | 1979 | 0.850 |
| 13 | Chile | 240 | 1.91 | 0.0494 | -195.7159 | 193905.2759 | 1981 | 0.950 |
| 14 | South Korea | 239 | 1.90 | 0.0445 | -176.1976 | 174445.0420 | 1980 | 0.847 |
| 15 | Venezuela | 222 | 1.77 | 0.0470 | -186.1174 | 184409.4761 | 1980 | 0.967 |
| 16 | Switzerland | 218 | 1.73 | 0.0418 | -165.5350 | 164003.5445 | 1980 | 0.899 |
| 17 | Italy | 204 | 1.62 | 0.0210 | -83.0401 | 82027.2685 | 1977 | 0.904 |
| 18 | Japan | 201 | 1.60 | 0.0200 | -78.8436 | 77835.8616 | 1971 | 0.857 |
| 19 | Sweden | 185 | 1.47 | 0.0530 | -210.2623 | 208543.3511 | 1984 | 0.828 |
| 20 | Ecuador | 181 | 1.44 | 0.0577 | -228.8310 | 226933.1359 | 1983 | 0.861 |
| 21 | China | 181 | 1.44 | 0.0631 | -250.4600 | 248419.1930 | 1985 | 0.888 |
| 22 | Australia | 144 | 1.15 | 0.0274 | -108.7253 | 107738.3919 | 1984 | 0.788 |
| 23 | Belgium | 120 | 0.95 | 0.0232 | -91.9566 | 91070.5231 | 1982 | 0.921 |
| 24 | Peru | 117 | 0.93 | 0.0168 | -66.3567 | 65715.6675 | 1975 | 0.861 |
| 25 | Costa Rica | 105 | 0.84 | 0.0187 | -74.0620 | 73359.4683 | 1980 | 0.824 |

Table 12: Main co-authorship networks of Cuba

| No. | Country | Co-publications (1973-2006) | % | *a* | *b* | *c* | $T_o$ | *R* |
|---|---|---|---|---|---|---|---|---|
| 1 | Spain | 1202 | 10.76 | 0.2641 | -1046.603 | 1036850.7810 | 1981 | 0.986 |
| 2 | Mexico | 880 | 7.88 | 0.1589 | -629.4325 | 623134.0570 | 1981 | 0.928 |
| 3 | Germany | 489 | 4.38 | 0.0530 | -209.7476 | 207591.3640 | 1979 | 0.881 |
| 4 | USA | 481 | 4.31 | 0.0647 | -256.1091 | 253430.2548 | 1979 | 0.955 |
| 5 | Brazil | 469 | 4.20 | 0.1132 | -448.8316 | 444774.5477 | 1982 | 0.967 |
| 6 | Italy | 353 | 3.16 | 0.0429 | -169.7399 | 167900.3852 | 1978 | 0.937 |
| 7 | France | 282 | 2.52 | 0.0228 | -90.0450 | 88784.0559 | 1975 | 0.938 |
| 8 | United Kingdom | 233 | 2.09 | 0.0523 | -207.1852 | 205352.3202 | 1981 | 0.922 |
| 9 | URSS/Russia | 210 | 1.88 | 0.1051 | -417.0247 | 413782.8451 | 1984 | 0.676 |
| 10 | Canada | 174 | 1.56 | 0.0344 | -136.5498 | 135314.5753 | 1985 | 0.949 |
| 11 | Belgium | 159 | 1.42 | 0.0500 | -198.3613 | 196708.9084 | 1984 | 0.925 |
| 12 | Denmark | 156 | 1.40 | 0.0500 | -198.2320 | 196577.6655 | 1982 | 0.924 |
| 13 | Argentina | 149 | 1.33 | 0.0406 | -161.2350 | 159887.8445 | 1986 | 0.929 |
| 14 | Sweden | 117 | 1.05 | 0.0126 | -50.2973 | 50232.7331 | 1996 | 0.710 |
| 15 | Chile | 112 | 1.00 | 0.0269 | -106.6625 | 105743.8905 | 1983 | 0.936 |
| 16 | Colombia | 92 | 0.82 | 0.0187 | -74.0620 | 73359.4683 | 1980 | 0.824 |
| 17 | Switzerland | 89 | 0.80 | 0.0059 | -23.2031 | 22856.4194 | 1966 | 0.824 |
| 18 | Venezuela | 82 | 0.73 | 0.0134 | -52.8952 | 52351.7649 | 1974 | 0.870 |
| 19 | Japan | 77 | 0.69 | 0.0174 | -68.8096 | 68201.0085 | 1977 | 0.841 |
| 20 | Netherlands | 55 | 0.49 | 0.0118 | -46.8303 | 46427.0429 | 1984 | 0.808 |
| 21 | Finland | 47 | 0.42 | 0.0068 | -26.9900 | 26703.8471 | 1985 | 0.768 |
| 22 | Portugal | 44 | 0.39 | 0.0026 | -10.5158 | 10432.4057 | 1997 | 0.677 |

Table 13: Main co-authorship networks of Peru

| No. | Country | Co-publications (1973-2006) | % | $a$ | $b$ | $c$ | $T_o$ | $R$ |
|---|---|---|---|---|---|---|---|---|
| 1 | USA | 1904 | 28.07 | 0.2328 | -922.7038 | 914228.5218 | 1982 | 0.954 |
| 2 | Spain | 449 | 6.62 | 0.0910 | -360.5324 | 357140.4285 | 1981 | 0.961 |
| 3 | United Kingdom | 355 | 5.23 | 0.0529 | -209.7974 | 207843.9493 | 1983 | 0.929 |
| 4 | Brazil | 341 | 5.03 | 0.0716 | -283.9686 | 281449.0659 | 1983 | 0.897 |
| 5 | France | 262 | 3.86 | 0.0354 | -140.2218 | 138874.3881 | 1981 | 0.911 |
| 6 | Germany | 216 | 3.18 | 0.0430 | -170.3971 | 168983.4590 | 1981 | 0.841 |
| 7 | Argentina | 190 | 2.80 | 0.0537 | -213.1090 | 211340.7053 | 1984 | 0.907 |
| 8 | Chile | 170 | 2.51 | 0.0373 | -148.0655 | 146822.6899 | 1985 | 0.872 |
| 9 | Mexico | 163 | 2.40 | 0.0416 | -164.9562 | 163543.3600 | 1983 | 0.909 |
| 10 | Canada | 160 | 2.36 | 0.0440 | -174.6967 | 173352.0021 | 1985 | 0.887 |
| 11 | Belgium | 143 | 2.11 | 0.0291 | -115.1882 | 114170.7013 | 1979 | 0.843 |
| 12 | Colombia | 117 | 1.72 | 0.0213 | -84.4426 | 83649.3938 | 1982 | 0.823 |
| 13 | Italy | 112 | 1.65 | 0.0239 | -94.8597 | 93979.6608 | 1985 | 0.869 |
| 14 | Switzerland | 104 | 1.53 | 0.0305 | -120.8426 | 119843.8870 | 1981 | 0.916 |
| 15 | Netherlands | 79 | 1.16 | 0.0177 | -70.0470 | 69466.0356 | 1979 | 0.882 |
| 16 | Ecuador | 71 | 1.05 | 0.0205 | -81.1051 | 80409.9235 | 1978 | 0.952 |
| 17 | Sweden | 71 | 1.05 | 0.0087 | -34.4280 | 34037.3707 | 1979 | 0.857 |
| 18 | Bolivia | 60 | 0.88 | 0.0167 | -66.1239 | 65566.2569 | 1980 | 0.886 |

Table 14: Main co-authorship networks of Uruguay

| No. | Country | Co-publications (1973-2006) | % | *a* | *b* | *c* | *T*$_o$ | *R* |
|---|---|---|---|---|---|---|---|---|
| 1 | USA | 931 | 14.94 | 0.1621 | -642.2450 | 635995.2735 | 1981 | 0.984 |
| 2 | Argentina | 625 | 10.03 | 0.1359 | -538.7078 | 533849.0381 | 1982 | 0.974 |
| 3 | Brazil | 492 | 7.90 | 0.1201 | -476.3842 | 472294.7166 | 1983 | 0.969 |
| 4 | Spain | 418 | 6.71 | 0.0966 | -383.0340 | 379542.6738 | 1983 | 0.976 |
| 5 | France | 349 | 5.60 | 0.0364 | -143.9800 | 142227.5940 | 1978 | 0.941 |
| 6 | Chile | 209 | 3.35 | 0.0456 | -180.5321 | 178886.6309 | 1980 | 0.961 |
| 7 | United Kingdom | 205 | 3.29 | 0.0269 | -106.4659 | 105292.5622 | 1979 | 0.944 |
| 8 | Sweden | 180 | 2.89 | 0.0156 | -61.6693 | 60802.2071 | 1977 | 0.882 |
| 9 | Germany | 177 | 2.84 | 0.0240 | -95.1656 | 94149.2565 | 1983 | 0.931 |
| 10 | Mexico | 156 | 2.50 | 0.0315 | -124.7898 | 123634.9285 | 1981 | 0.950 |
| 11 | Italy | 144 | 2.31 | 0.0273 | -108.2157 | 107165.5126 | 1982 | 0.938 |
| 12 | Canada | 95 | 1.52 | 0.0141 | -55.9405 | 55355.3577 | 1984 | 0.923 |
| 13 | Japan | 68 | 1.09 | 0.0118 | -46.6062 | 46138.9291 | 1975 | 0.864 |
| 14 | Venezuela | 67 | 1.08 | 0.0120 | -47.3477 | 46889.7717 | 1973 | 0.829 |
| 15 | Australia | 64 | 1.03 | 0.0173 | -68.6734 | 68087.2582 | 1985 | 0.871 |
| 16 | Netherlands | 51 | 0.82 | 0.0082 | -32.3819 | 32044.9298 | 1975 | 0.878 |
| 17 | Colombia | 44 | 0.71 | 0.0116 | -46.1537 | 45773.6101 | 1989 | 0.829 |
| 18 | Belgium | 43 | 0.69 | 0.0030 | -11.7190 | 11535.7625 | 1983 | 0.784 |
| 19 | Switzerland | 38 | 0.61 | 0.0084 | -33.3342 | 35025.7405 | 1984 | 0.846 |

Table 15: Main co-authorship networks of Costa Rica

| No. | Country | Co-publications (1973-2006) | % | $a$ | $b$ | $C$ | $T_o$ | $R$ |
|-----|---------|------|-------|--------|-----------|-------------|------|-------|
| 1 | USA | 1718 | 27.79 | 0.1349 | -533.1565 | 526963.1168 | 1976 | 0.977 |
| 2 | Mexico | 305 | 4.93 | 0.0437 | -173.0505 | 171810.1219 | 1980 | 0.878 |
| 3 | Germany | 261 | 4.22 | 0.0304 | -120.1101 | 118832.1219 | 1975 | 0.898 |
| 4 | France | 256 | 4.14 | 0.0387 | -153.1364 | 151558.9637 | 1979 | 0.929 |
| 5 | Brazil | 205 | 3.32 | 0.0471 | -186.9644 | 185375.6686 | 1985 | 0.895 |
| 6 | United Kingdom | 196 | 3.17 | 0.0219 | -86.5575 | 85591.5007 | 1976 | 0.896 |
| 7 | Netherlands | 190 | 3.07 | 0.0247 | -97.5604 | 96457.1388 | 1975 | 0.911 |
| 8 | Spain | 176 | 2.85 | 0.0331 | -131.1052 | 139837.3730 | 1980 | 0.948 |
| 9 | Canada | 175 | 2.83 | 0.0227 | -89.7466 | 88785.8550 | 1977 | 0.931 |
| 10 | Sweden | 166 | 2.69 | 0.0051 | -19.7191 | 19171.6257 | 1933 | 0.830 |
| 11 | Panamá | 107 | 1.73 | 0.0071 | -28.1420 | 27769.2339 | 1982 | 0.751 |
| 12 | Colombia | 105 | 1.70 | 0.0187 | -74.0620 | 73359.4683 | 1980 | 0.824 |
| 13 | Argentina | 85 | 1.38 | 0.0222 | -88.0110 | 87251.5958 | 1982 | 0.842 |
| 14 | Italy | 76 | 1.23 | 0.0124 | -48.9939 | 48487.4417 | 1976 | 0.856 |
| 15 | Israel | 59 | 0.95 | 0.0077 | -30.5322 | 30192.2090 | 1983 | 0.733 |
| 16 | Japan | 55 | 0.89 | 0.0051 | -20.0060 | 19770.6469 | 1961 | 0.814 |
| 17 | Chile | 53 | 0.86 | 0.0184 | -73.1392 | 72561.8009 | 1987 | 0.857 |
| 18 | Guatemala | 51 | 0.83 | 0.0074 | -29.1484 | 28887.8117 | 1979 | 0.779 |
| 19 | Nicaragua | 50 | 0.81 | 0.0149 | -58.9963 | 58504.2246 | 1980 | 0.844 |